\documentclass[aps,preprint,tightenlines,superscriptaddress,nofootinbib]{revtex4-1}

\usepackage[utf8]{inputenc}
\usepackage{amsmath,amssymb}
\usepackage{braket}
\usepackage[colorlinks=true,linkcolor=blue,urlcolor=blue,citecolor=blue,anchorcolor=blue]{hyperref}
\usepackage[capitalise]{cleveref}
\Crefname{section}{Sec.}{Secs.}
\crefrangelabelformat{equation}{(#3#1#4--#5#2#6)}
\usepackage{tikz}
\usepackage{multirow}
\usepackage{amsfonts}
\usepackage{ulem}
\usepackage{enumitem}

\begin{document}

\title{Hamiltonian effective field theory in elongated or moving finite volume}
\newcommand{\UCAS}{School of Physical Sciences, University of Chinese Academy of Sciences (UCAS), Beijing 100049, China}
\newcommand{\CSSM}{Special Research Centre for the Subatomic Structure
of Matter (CSSM),\\Department of Physics, University of
Adelaide, Adelaide, South Australia 5005, Australia}
\newcommand{\CoEPP}{ARC Centre of Excellence for Particle Physics at
the Terascale (CoEPP),\\Department of Physics, University
of Adelaide, Adelaide, South Australia 5005, Australia}
\author{Yan Li}
\affiliation{\UCAS}
\author{Jia-Jun Wu}
\affiliation{\UCAS}
\author{Derek B. Leinweber}
\affiliation{\CSSM}
\author{Anthony W. Thomas}
\affiliation{\CSSM}
\affiliation{\CoEPP}

\begin{abstract}
We extend previous work concerning rest-frame partial-wave mixing in Hamiltonian effective field theory to both elongated and moving systems, where two particles are in a periodic elongated cube or have nonzero total momentum, respectively. We also consider the combination of the two systems when directions of the elongation and the moving momentum are aligned.
This extension should also be applicable in any Hamiltonian formalism.
As a demonstration, we analyze lattice QCD results for the spectrum of an isospin-2 $\pi\pi$ scattering system and determine the $s$, $d$, and $g$ partial-wave scattering information.
The inclusion of lattice simulation results from moving frames significantly improves the
uncertainty in the scattering information.
\end{abstract}

\maketitle
\newpage
\section{Introduction}
Lattice simulations of relativistic quantum-field theories are performed in a Euclidean
four-dimensional finite volume.  Scattering states are contained in the finite box with discretized energy levels.
Understanding the relationship between these finite-volume energy levels and experimental scattering observables such as the phase shift and inelasticity is of significance.
For the case of elastic two-body scattering in the rest frame, L\"{u}scher
\cite{Luscher:1985dn,Luscher:1986pf,Luscher:1990ux} found a model-independent formula which is now
known as L\"{u}scher's formula.

An equivalent approach is provided in Hamiltonian effective field theory (HEFT)
\cite{Hall:2013qba,Hall:2014uca,Wu:2014vma,Liu:2015ktc,Liu:2016uzk,Liu:2016wxq,Wu:2016ixr,Wu:2017qve,Li:2019qvh},
a Hamiltonian extension of chiral effective field theory. 
In the standard approach, a Hamiltonian which respects the constraints of chiral effective field
theory is fit to the finite-volume energy spectrum of lattice field theory and the infinite-volume
scattering observables are obtained from the constrained Hamiltonian. The approach bridges
finite-volume lattice field theory and experimental observables while providing insight into the
composition of the scattering states in terms of noninteracting multiparticle basis states.

Different partial waves are mixed in the finite volume as a result of broken spherical symmetry.
This mixing complicates the construction of the Hamiltonian matrix. For example, its incorporation
significantly increases the dimension of the matrix.
A recent work \cite{Li:2019qvh} established a formalism for disentangling partial-wave mixing and maximally reducing the dimension of the Hamiltonian matrix in the finite volume via an optimal set of rest-frame basis states.
In this work, we will generalize this formalism to both elongated and moving systems with
nontrivial total momentum. We will also consider the combination of these two systems when the
direction of elongation and that of the moving momentum are aligned. 

The L\"{u}scher formula has already been extended and applied to the case of rectangular cuboid boxes \cite{Feng:2004ua,Li:2003jn,Li:2007ey,Lee:2017igf,Li:2018tzx,Meng:2009qt,Pelissier:2012pi,Guo:2016zos,Guo:2018zss,Culver:2019qtx,Culver:2019vvu}.
This work will first consider a more general case where the box is allowed to be a general
parallelepiped, as illustrated in \cref{fig:elm-par}. We then focus on a special class of the
parallelepiped termed an elongated cube. 

The L\"{u}scher formula has also been extended to moving systems~\cite{Rummukainen:1995vs,Kim:2005gf,Gockeler:2012yj,Davoudi:2011md,Fu:2011xz,Leskovec:2012gb}.
To realize the extension in a Hamiltonian formalism, one needs a Hamiltonian making contact with
both the infinite-volume scattering observables parametrized in the rest frame and the
finite-volume spectrum in the moving frame. 
This can be achieved within the formalism proposed in Refs.~\cite{Li:2021General,Wu:2015evh}.
In that formalism, different forms of the moving-frame L\"{u}scher formula are unified as different
momentum transformations. Furthermore it leads to a new momentum transformation which is not only
useful in the Hamiltonian formalism, but can also be used in the finite-volume three-particle
quantization condition \cite{Blanton:2020gha}. 

The symmetry in a moving frame is quite compatible with a cube elongated in the same direction as
the nonzero total momentum. 
This case will be termed the elongated moving system, and disentangling partial-wave mixing in the
elongated moving system is the main concern of this work. 
We will also demonstrate how the formalism works by analyzing lattice QCD results from
Ref.~\cite{Dudek:2012gj} for the spectrum of an isospin-2 $\pi\pi$ scattering system.
As also noted in Ref.~\cite{Li:2019qvh}, the discussion in this work should apply not only in HEFT,
but also in any Hamiltonian formalism, e.g., the harmonic oscillator basis effective theory
\cite{McElvain:2017Harmonic,McElvain:2019ltw,Drischler:2019xuo}.

There are also many other extensions of the L\"{u}scher formula, including the multichannel case \cite{Bernard:2010fp,Guo:2012hv,He:2005ey,Hu:2016shf,Lage:2009zv,Li:2012bi}, nonzero spins \cite{Beane:2003da,Beane:2003yx,Briceno:2013bda,Meng:2003gm}, twisted-boundary conditions \cite{Bedaque:2004ax,Bedaque:2004kc,deDivitiis:2004kq,Sachrajda:2004mi} and the multibody case \cite{Polejaeva:2012ut,Briceno:2012rv,Hansen:2014eka,Hansen:2015zga,Briceno:2017tce,Hammer:2017kms,Hammer:2017uqm,Hansen:2017mnd,Mai:2017bge,Doring:2018xxx,Guo:2018ibd,Mai:2018djl,Meng:2017jgx,Bulava:2019kbi,Hansen:2019nir,Jackura:2019bmu,Pang:2019dfe,Romero-Lopez:2019qrt,Blanton:2019vdk,Blanton:2020gha,Hansen:2020otl}.
With the exception of the many-body case, these extensions should be easily realized in the Hamiltonian formalism using the results of Ref.~\cite{Li:2019qvh} and this paper.
In addition, there have also been studies concerning finite-spacing effects in the Hamiltonian formalism, e.g., Ref.~\cite{Korber:2019cuq}.

This paper is organized as follows.
In \cref{sec:mtems}, the finite-volume Hamiltonian in the elongated moving system is established.
\cref{sec:pwmems} accommodates partial-wave mixing in the elongated moving system using the
formalism developed in Ref.~\cite{Li:2019qvh}. 
\cref{sec:eis} demonstrates how this formalism works by analyzing lattice QCD results for isospin-2 $\pi\pi$ scattering \cite{Dudek:2012gj}.
Finally, the results are summarized in \cref{sec:sum}.

\section{Hamiltonian in elongated moving finite volume}\label{sec:mtems}
\subsection{Parallelepiped and elongated cube}\label{sec:PEC}
\begin{figure}[tbp]
    \centering
    \includegraphics[width=15cm]{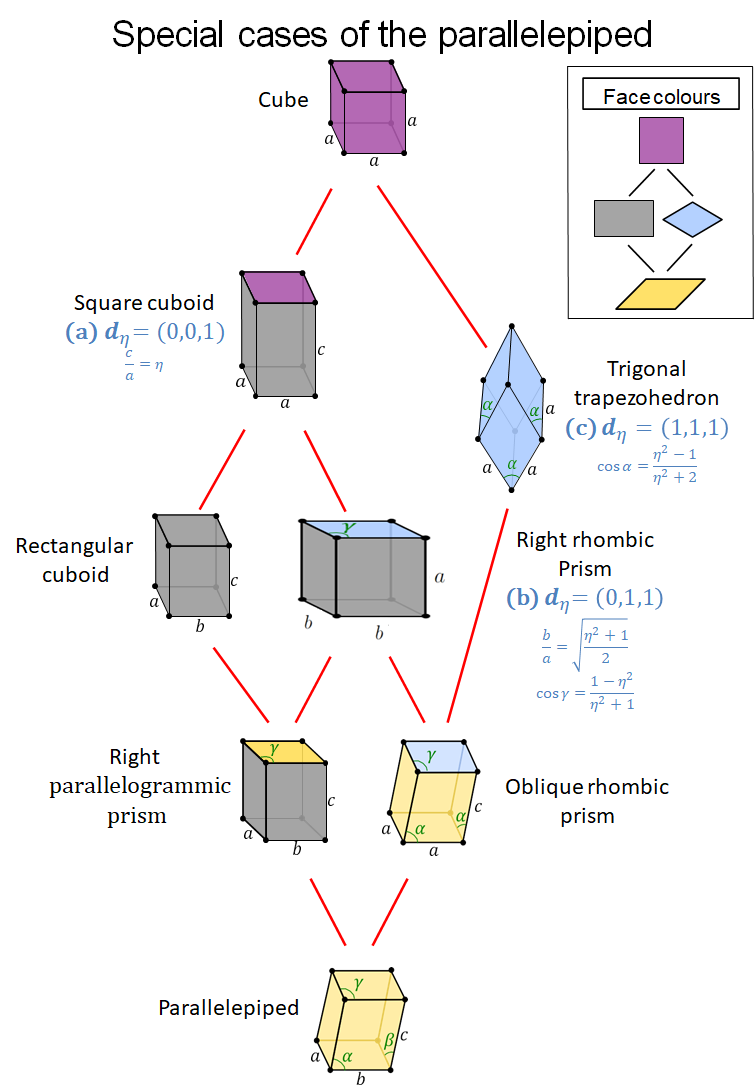}
    \caption{Special cases of the parallelepiped (taken from Ref.~\cite{:2019Parallelepiped} with
      slight modifications). The elongated cubes discussed in this work are labelled with the
      elongated vector $\mathbf{d}_\eta$.
    }
\label{fig:elm-par} 
\end{figure}

Normally, the system under consideration in lattice field theory
simulations is a periodic cube. However, there are good reasons to
also consider asymmetric boxes, where longer dimensions provide access
to smaller nontrivial momenta \cite{Leinweber:1990dv}. For example,
Ref.~\cite{Feng:2004ua} studied a rectangular cuboid (including the
square cuboid as a special case). In general, the box can be a
parallelepiped as shown in \cref{fig:elm-par}.

If we choose one of the vertices of the parallelepiped as the origin, the parallelepiped can be specified by the three vectors $\mathbf{a},\mathbf{b},\mathbf{c}$ corresponding to the edges connected to the origin. The three vectors specify a matrix
\begin{align}
    M =
    \begin{bmatrix}
        a_1 & b_1 & c_1 \\
        a_2 & b_2 & c_2 \\
        a_3 & b_3 & c_3 \\
    \end{bmatrix} \, ,
\end{align}
which sends the three unit vectors of the coordinate axes to $\mathbf{a},\mathbf{b},$ and $\mathbf{c}$ respectively, where the subscripts denote the coordinate components of the vectors.
To restrict the range of $\mathbf{x}^*$ within the parallelepiped, one can define $\mathbf{x^*}$ via $x^*_i=\sum_j M_{ij}\, x_{j}$ and constrain $\mathbf{x}$ within the unit cube.
Correspondingly, when imposing the periodic boundary condition, the momentum $\mathbf{k}^*$ should be discretized as
\begin{align}
    k^*_i  = \sum_j M^{-1}_{ij}\,\frac{2\pi}{L} n_j, \quad n_j \in \mathbb{Z} \,,
\end{align}
where we replace the unit cube with a cube of edge length $L$, as is standard in lattice field
theory. 

Here, we do not consider all the cases in \cref{fig:elm-par} since in most cases the symmetries are broken too much.
When we study moving frames in the following sections, we will find that the symmetry of a cube elongated in the same direction as the moving momentum is quite compatible with the moving effects.
So in this paper, we define the elongated cube as a cube elongated in a specific direction $\hat{\mathbf{d}}_\eta$ with a magnitude $\eta$, and we consider three $\mathbf{d}_\eta$ as follows:

\begin{enumerate}[label=(\alph*)]
    \item $\mathbf{d}_\eta=(0,0,1)$, $M=
    \begin{bmatrix}
        1 & & \\
         & 1 & \\
         & & \eta \\
    \end{bmatrix}$, corresponding to the square cuboid (already covered in Ref.~\cite{Feng:2004ua}), labelled as (a) in \cref{fig:elm-par}.
    \item $\mathbf{d}_\eta=(0,1,1)$, $M=
    \begin{bmatrix}
        1 & 0 & 0 \\
        0 & \frac{\eta+1}{2} & \frac{\eta-1}{2} \\
        0 & \frac{\eta-1}{2} & \frac{\eta+1}{2} \\
    \end{bmatrix}$, corresponding to the right rhombic prism, labelled as (b) in \cref{fig:elm-par}.
    We note that not all right rhombic prisms are included in this scenario, since $\frac{b}{a}=\sqrt{\frac{\eta^2+1}{2}}$ and $\cos \gamma=\frac{1-\eta^2}{\eta^2+1}$ are both determined by $\eta$.
        The general right rhombic prism corresponds to $M=
        \begin{bmatrix}
            \eta_x & 0 & 0 \\
            0 & \frac{\eta+1}{2} & \frac{\eta-1}{2} \\
            0 & \frac{\eta-1}{2} & \frac{\eta+1}{2} \\
        \end{bmatrix}$, which is an elongated cube only when $\eta_x=1$.
    \item $\mathbf{d}_\eta=(1,1,1)$, $M=
    \begin{bmatrix}
        \frac{\eta+2}{3} & \frac{\eta-1}{3} & \frac{\eta-1}{3} \\
        \frac{\eta-1}{3} & \frac{\eta+2}{3} & \frac{\eta-1}{3} \\
        \frac{\eta-1}{3} & \frac{\eta-1}{3} & \frac{\eta+2}{3} \\
    \end{bmatrix}$, corresponding to the trigonal trapezohedron, labelled as (c) in \cref{fig:elm-par}.
\end{enumerate}

We note that the overall factors of $\mathbf{d}_\eta$ are not important, and are taken as presented
for further convenience.
In the elongated cube, the momentum
is discretized as
\begin{align}\label{eq:ats}
    \mathbf{k}^*=\frac{2\pi}{L} \left( \mathbf{n}_{\perp} + \frac{1}{\eta}\mathbf{n}_{\parallel} \right) \,,\quad \mathbf{n} \in \mathbb{Z}^3 \,, 
\end{align}
where the $\perp$ and $\parallel$ components of a vector are defined through
\begin{align}
    \mathbf{n}_{\parallel} := \frac{\mathbf{n}\cdot\mathbf{d}_\eta}{|\mathbf{d}_\eta|^2} \mathbf{d_\eta} \,, \quad \mathbf{n}_{\perp} := \mathbf{n} - \mathbf{n}_{\parallel} \,.
\end{align}

To be more concrete, for an infinite-volume Hamiltonian
\begin{align}
    H &= \int \frac{d^3 \mathbf{k}^*}{(2\pi)^3}\, h(k^*)\, \ket{\mathbf{k}^*}\bra{\mathbf{k}^*} +
    \int \frac{d^3 \mathbf{k}'^*}{(2\pi)^3}\frac{d^3 \mathbf{k}^*}{(2\pi)^3} \,
    V(\mathbf{k}'^*,\mathbf{k}^*)\,  \ket{\mathbf{k}'^*}\bra{\mathbf{k}^*} \,,
\end{align}
where $h$ and $V$ denote the kinetic and potential energy respectively, and the state $\ket{\mathbf{k}^*}$ is normalized as
\begin{align}
    \braket{\mathbf{k}'^*|\mathbf{k}^*} = (2\pi)^3\, \delta^3(\mathbf{k}'^*-\mathbf{k}^*) \,,
\end{align}
to put it in an elongated cube, we need the discretization
\begin{align}
    \int \frac{d^3 \mathbf{k}^*}{(2\pi)^3} \to \eta^{-1}L^{-3}\sum_{\mathbf{k}^*=\frac{2\pi}{L} \left( \mathbf{n}_{\perp} + \frac{1}{\eta}\mathbf{n}_{\parallel} \right),~\mathbf{n}\in\mathbb{Z}^3}\,,
\end{align}
and
\begin{align}
    \ket{\mathbf{k}^*} \to \eta^{\frac{1}{2}}L^{\frac{3}{2}}\,\ket{\mathbf{n}}
\end{align}
so that the basis $\ket{\mathbf{n}}$ is orthonormal
\begin{align}
    \braket{\mathbf{n}'|\mathbf{n}}=\delta_{\mathbf{n}'\, \mathbf{n}}\,.
\end{align}
Finally, the Hamiltonian in a finite elongated cube is
\begin{align}\label{eq:fvhe}
    H_L &= \sum_{\mathbf{n}} \, h(k^*(\mathbf{n}))\, \ket{\mathbf{n}}\bra{\mathbf{n}} +
    \sum_{\mathbf{n}',\mathbf{n}} \eta^{-1}\, L^{-3}\,
    V(\mathbf{k}^*(\mathbf{n}'),\mathbf{k}^*(\mathbf{n}))\, \ket{\mathbf{n}'}\bra{\mathbf{n}} \,.  
\end{align}
We note the Hamiltonian \cref{eq:fvhe} applies to any elongated cubes, including those of the three scenarios (a), (b) and (c) introduced above \cref{eq:ats}.

We also note, that the box studied in Ref.~\cite{Feng:2004ua} corresponds to
\begin{align}\label{eq:elm-fengM}
    M = 
    \begin{bmatrix}
        \eta_1 & & \\
         & \eta_2 & \\
         & & \eta_3 \\
    \end{bmatrix}\,,
\end{align}
where an overall factor can be absorbed into $L$. This box corresponds to the rectangular cuboid, which is not completely equivalent to the elongated cube, since it needs elongation in more than one direction.
However in some cases, $\eta_1=\eta_2=1$ for instance, the box is an elongated cube with $\mathbf{d}_\eta=(0,0,1)$.

\subsection{Moving system}\label{sec:MS}
Since the infinite-volume potential and scattering observables are most easily parametrized in the rest frame, we need a Hamiltonian that can produce the moving-frame spectrum while still written in terms of the rest-frame potential.

As suggested in Refs.~\cite{Li:2021General,Wu:2015evh}, one can introduce a momentum transformation $\mathbf{k}^*\to\mathbf{k}$ to the infinite-volume Hamiltonian
\begin{align}
    H &= \int \frac{d^3 \mathbf{k}^*}{(2\pi)^3} \, h(k^*)\, \ket{\mathbf{k}^*}\bra{\mathbf{k}^*} + \int \frac{d^3 \mathbf{k}'^*}{(2\pi)^3}\frac{d^3 \mathbf{k}^*}{(2\pi)^3} \, V(\mathbf{k}'^*,\mathbf{k}^*)\, \ket{\mathbf{k}'^*}\bra{\mathbf{k}^*} \nonumber\\
    &= \int \frac{d^3 \mathbf{k}}{(2\pi)^3}\,\mathcal{J}(\mathbf{k})~h(k^*(\mathbf{k}))\, \ket{\mathbf{k}^*(\mathbf{k})}\bra{\mathbf{k}^*(\mathbf{k})} \nonumber\\
    &+ \int \frac{d^3 \mathbf{k}'}{(2\pi)^3}\,\mathcal{J}(\mathbf{k}')~\frac{d^3 \mathbf{k}}{(2\pi)^3}\,\mathcal{J}(\mathbf{k})~V(\mathbf{k}^*(\mathbf{k}'),\mathbf{k}^*(\mathbf{k}))\,\ket{\mathbf{k}^*(\mathbf{k}')}\bra{\mathbf{k}^*(\mathbf{k})} \,,
\end{align}
where $\mathcal{J}$ denotes the Jacobian of the transformation.
Then one can define
\begin{align}
    \ket{\mathbf{k}} := \mathcal{J}^{\frac{1}{2}}(\mathbf{k}) \,\ket{\mathbf{k}^*(\mathbf{k})} \,,
\end{align}
such that
\begin{align}
    \braket{\mathbf{k}'|\mathbf{k}} = \mathcal{J}(\mathbf{k}) (2\pi)^3\delta^3(\mathbf{k}'^*-\mathbf{k}^*) = (2\pi)^3\delta^3(\mathbf{k}'-\mathbf{k}) \,,
\end{align}
and the infinite-volume Hamiltonian will be
\begin{align}\label{eq:elm-ifht}
    H = \int \frac{d^3 \mathbf{k}}{(2\pi)^3} \, h(k^*(\mathbf{k}))\,
    \ket{\mathbf{k}}\bra{\mathbf{k}} + \int \frac{d^3 \mathbf{k}'}{(2\pi)^3}\frac{d^3
      \mathbf{k}}{(2\pi)^3} \left [\mathcal{J}^{\frac{1}{2}}(\mathbf{k}')\, V(\mathbf{k}^*(\mathbf{k}'),\mathbf{k}^*(\mathbf{k}))\,\mathcal{J}^{\frac{1}{2}}(\mathbf{k})\right]\,\ket{\mathbf{k}'}\bra{\mathbf{k}} \,.
\end{align}
Now if one discretizes $\mathbf{k}$ instead of $\mathbf{k}^*$, one gets a different finite-volume Hamiltonian
\begin{align}\label{eq:dfvHMoving}
    &H_L = \sum_{\mathbf{n}} \, h(k^*(\mathbf{n}))\, \ket{\mathbf{n}}\bra{\mathbf{n}} + \sum_{\mathbf{n}',\mathbf{n}} L^{-3} \,\tilde{V}(\mathbf{k}(\mathbf{n}'),\mathbf{k}(\mathbf{n}))\, \ket{\mathbf{n}'}\bra{\mathbf{n}} \nonumber\\
    &\tilde{V}(\mathbf{k}',\mathbf{k}) = \left [\mathcal{J}^{\frac{1}{2}}(\mathbf{k}')\, V(\mathbf{k}'^*,\mathbf{k}^*)\,\mathcal{J}^{\frac{1}{2}}(\mathbf{k})\right]\,,\qquad\mathbf{k}=\frac{2\pi}{L}\mathbf{n} \,.
\end{align}

Ref.~\cite{Li:2021General} proposed a number of general momentum transformations for the moving system, and proved that those transformations can keep the relationship between the infinite-volume phase shifts and finite-volume spectrum up to exponentially suppressed corrections.
That paper also studied three typical transformations.
While two of the three have been used in many previous works \cite{Rummukainen:1995vs,Kim:2005gf,Gockeler:2012yj}, they introduce additional energy dependence. 
The third one (labelled as scheme C in Ref.~\cite{Li:2021General}) does not have such problems and hence is suitable here. It reads
\begin{align}\label{eq:mmt}
    &\mathbf{k}^* = \mathbf{k}_{\perp} + \gamma \left( \mathbf{k}_{\parallel} - \frac{\omega_1(\mathbf{k})}{\omega_1(\mathbf{k})+\omega_2(\mathbf{P}-\mathbf{k})} \mathbf{P} \right) \,, \nonumber\\
    &\gamma = \frac{\omega_1(\mathbf{k})+\omega_2(\mathbf{P}-\mathbf{k})}{\sqrt{\left( \omega_1(\mathbf{k})+\omega_2(\mathbf{P}-\mathbf{k}) \right)^2-\mathbf{P}^2}} \,,
\end{align}
where $\omega_i(\mathbf{k})=\sqrt{k^2+m_i^2}$ and $\mathbf{P}$ is the total momentum of the moving system, and the corresponding Jacobian is
\begin{align}\label{eq:jacob}
    &\mathcal{J}(\mathbf{k})=\frac{\omega_1(\mathbf{k})+\omega_2(\mathbf{P}-\mathbf{k})}{\omega_1(\mathbf{k})\,\omega_2(\mathbf{P}-\mathbf{k})}\bigg{/}\frac{\omega_1(\mathbf{k}^*)+\omega_2(\mathbf{k}^*)}{\omega_1(\mathbf{k}^*)\,\omega_2(\mathbf{k}^*)} \,.
\end{align}
In the finite cube, the total momentum can only take discrete values as $\mathbf{P}=\frac{2\pi}{L} \mathbf{d}_\gamma$ with $\mathbf{d}_\gamma$ an integer vector.

\subsection{Elongated moving system}\label{sec:EMS}
Both the elongated and moving systems have smaller finite-volume symmetry groups than the rest-frame cube, and combining them will normally give a much smaller one. 
However, if the elongated direction and the moving direction are the same, their combination will not see a large reduction in symmetry.
We will call this combination the elongated moving system. The corresponding finite-volume Hamiltonian is obtained by combining \cref{eq:fvhe,eq:dfvHMoving}, which reads
\begin{align}\label{eq:dfvH}
    &H_L = \sum_{\mathbf{n}} \, h(k^*(\mathbf{n}))\, \ket{\mathbf{n}}\bra{\mathbf{n}} + \sum_{\mathbf{n}',\mathbf{n}} \eta^{-1}L^{-3} \,\tilde{V}(\mathbf{k}(\mathbf{n}'),\mathbf{k}(\mathbf{n}))\, \ket{\mathbf{n}'}\bra{\mathbf{n}} \nonumber\\
    &\tilde{V}(\mathbf{k}',\mathbf{k}) = \mathcal{J}^{\frac{1}{2}}(\mathbf{k}')\, V(\mathbf{k}'^*,\mathbf{k}^*)\,\mathcal{J}^{\frac{1}{2}}(\mathbf{k})\,,\qquad\mathbf{k}=\frac{2\pi}{L} \left( \mathbf{n}_{\perp} + \frac{1}{\eta}\mathbf{n}_{\parallel} \right) \,,
\end{align}
where the momentum transformation $\mathbf{k}^*\to \mathbf{k}$ and the corresponding Jacobian is  the same as \cref{eq:mmt,eq:jacob} for the moving system, except that the total momentum $\mathbf{P}$ should now be
\begin{align}\label{eq:tmp}
    \mathbf{P}=\frac{1}{\eta} \frac{2\pi}{L} \mathbf{d}_\gamma \, ,
\end{align}
noting either $\mathbf{d}_\eta = \mathbf{d}_\gamma$ or $\eta = 1$ for $\mathbf{P}\ne \mathbf{0}$ in this ``elongated moving system''.

Now \cref{eq:mmt,eq:jacob,eq:dfvH,eq:tmp} are all the ingredients needed to write down the elongated moving Hamiltonian.
The eigenvalues of this Hamiltonian are rest-frame energies $E^*_n$ related to moving-frame energies $E_n$ via
\begin{align}
    E_n=\sqrt{E^{*\,2}_n+\mathbf{P}^2} \,.
\end{align}

Since the kinetic energy $h(k^*)$ depends only on the length of $\mathbf{k}^*$, one can separate the kinetic term as
\begin{align}
    \sum_{\mathbf{n}} \, h(k^*)\, \ket{\mathbf{n}}\bra{\mathbf{n}} = \sum_{e_n} \, h(k^*)\, \sum_{\hat{e}_n} \ket{\mathbf{n}}\bra{\mathbf{n}} \,,
\end{align}
so that $k^*$ depends only on $e_n$ (and $\sum_{\hat{e}_n}$ is simply defined to sum over all the $\mathbf{n}$ with the same $e_n$).
In other words, $e_n$ denotes a degenerate shell of the basis-state Hamiltonian, $H_0$,
describing the energies of the noninteracting states.
Note, there can be several values for $e_n$ related to vectors $\mathbf{n}$, $\mathbf{d}_\eta$ and
$\mathbf{d}_\gamma$ all providing the same degenerate value for $h(k^*)$.

It is interesting to consider how the elongated moving system reduces to more simple cases for
certain values of $\mathbf{d}_\eta$, $\eta$ and $\mathbf{d}_\gamma$.  When the masses of the two
particles are the same, we also have
\begin{align}\label{eq:ndn}
    \mathbf{n}\to\mathbf{d}_\gamma-\mathbf{n} \quad\Rightarrow\quad \mathbf{k}^*\to -\mathbf{k}^* \quad\Rightarrow\quad h(k^*) \text{ invariant.}
\end{align}
Thus our discussion for the degenerate shells splits into four cases as listed in \cref{tab:defc}. 
There we introduce an elongated moving vector $\mathbf{d}\neq\mathbf{0}$, since either we can set
$\mathbf{d}_\eta=\mathbf{d}_\gamma$ or choose one of the two vectors to vanish.
In \cref{tab:defc}, case A refers to the standard unelongated and 
rest-frame system. $e_n$ can be simply chosen as $\mathbf{n}^2$.
Case B refers to the elongated or unelongated moving-frame system with two particles of different mass. As suggested by \cref{eq:mmt}, $h(k^*)$ now depends on $\mathbf{n}^2$, $(\mathbf{d-n})^2$, $\mathbf{n}_\parallel^2$, and $\mathbf{n}_\perp^2$. Because both $\mathbf{n}_\parallel^2$ and $\mathbf{n}_\perp^2$ can be reexpressed in terms of $\mathbf{n}^2$ and $(\mathbf{d-n})^2$, $e_n$ can be chosen as $\left(\mathbf{n}^2,(\mathbf{d}-\mathbf{n})^2\right)$, or $(\mathbf{n}^2,\mathbf{n\cdot d})$ equivalently.
Case C1 refers to the elongated rest-frame system. $h(k^*)$ now depends on $\mathbf{n}^2$, $\mathbf{n}_\parallel^2\propto |\mathbf{n\cdot d}|^2$, and $\mathbf{n}_\perp^2=\mathbf{n}^2-\mathbf{n}_\parallel^2$. Thus $e_n$ can be chosen as $\left(\mathbf{n}^2,|\mathbf{n\cdot d}|\right)$.
Case C2 refers to the elongated or unelongated moving-frame system with two particles of the same mass. In contrast to case B, $\mathbf{n}$ and $\mathbf{d-n}$ are on the same shell as indicated by \cref{eq:ndn}. As $\left(\mathbf{n}^2,(\mathbf{d-n})^2\right)$ and $\left((\mathbf{d-n})^2,\mathbf{n}^2\right)$ denote the same shell, $e_n$ can be chosen as an unordered pair $\left\{\mathbf{n}^2,(\mathbf{d-n})^2\right\}$.

\begin{table}[tbp]
    \centering
    \caption{Four different cases for the degenerate shells.
    The $\left\{\mathbf{n}^2,(\mathbf{d}-\mathbf{n})^2\right\}$ in the C2 row is an unordered pair.
    }
    \label{tab:defc}
    \renewcommand\arraystretch{1.5}
    \begin{ruledtabular}
    \begin{tabular}{cccccc}
        Case & $\mathbf{d}_\eta$ & $\eta$ & $\mathbf{d}_\gamma$ & $m_1=m_2$? & $e_n$ \\ \hline
        A & Any & $=1$ & $\mathbf{0}$ & Any & $\mathbf{n}^2$ \\
        B & $\mathbf{d}\neq\mathbf{0}$ & Any & $\mathbf{d}\neq\mathbf{0}$ & No &  $\left(\mathbf{n}^2,(\mathbf{d}-\mathbf{n})^2\right)$ or $\left(\mathbf{n}^2,\mathbf{n}\cdot\mathbf{d}\right)$ \\
        C1 & $\mathbf{d}\neq\mathbf{0}$ & $\neq 1$ & $\mathbf{0}$ & Any & $\left(\mathbf{n}^2,|\mathbf{n}\cdot\mathbf{d}|\right)$ \\
        C2 & $\mathbf{d}\neq\mathbf{0}$ & Any & $\mathbf{d}\neq\mathbf{0}$ & Yes & $\left\{\mathbf{n}^2,(\mathbf{d}-\mathbf{n})^2\right\}$ \\
    \end{tabular}
    \end{ruledtabular}
\end{table}
%

\section{Partial-wave mixing in an elongated moving system}\label{sec:pwmems}

Spherical symmetry allows the following partial-wave expansion
\begin{align}\label{eq:pwe}
    &V(\mathbf{k}'^*,\mathbf{k}^*) = \sum_{l} v_l(k'^*,k^*)\,\sum_m Y_{lm}(\hat{\mathbf{k}}'^*)\,Y_{lm}^*(\hat{\mathbf{k}}^*) \,,
\end{align}
where $Y_{lm}(\hat{\mathbf{k}}^*)$ are the usual spherical harmonics, as shown in \cref{eq:APPB1}, and its variables are the direction angles $(\theta^*,\phi^*)$ of the vector $\mathbf{k}^*$.
Different partial waves are decoupled under this potential in the infinite volume.
In the finite volume, partial wave numbers $(l,m)$ are no longer good quantum numbers, and the partial wave potentials $v_l$ with different $l$ are coupled together in the determination of finite-volume spectra.
This phenomena, called partial-wave mixing, complicates the structure of the Hamiltonian \cref{eq:dfvH}.
In the standard case (case A in \cref{tab:defc}), Ref.~\cite{Li:2019qvh} proposed a method that provides an optimal set of basis states maximally reducing the dimension of the Hamiltonian. In this section, we will generalize that method to more general cases.

The spherical symmetry group SO(3) is broken into one of its subgroups $G$ in the finite volume.
In the standard case A, $G$ is the octahedral group O. 
In other cases in \cref{tab:defc}, $G$ is smaller, and turns out to be also a subgroup of a two-dimensional rotation group O(2) (or O(2)$\times$C$_2$ in case C), where the rotation axis of this O(2) should be the same as the elongated moving vector $\mathbf{d}$. 
This prefers the partial wave expansion \cref{eq:pwe} expanded in a coordinate system different from that of the discretized momentum.
\cref{app:SGR} discusses how the coordinate system is chosen (results are summarized in \cref{tab:gfv-xyz}).

\begin{table}[tbp]
    \centering
    \caption{Some definitions for different cases. $\alpha_\infty$ denotes the index of the vector of the irreducible representation $\Gamma_\infty$ of the group $G_\infty$.}\label{tab:elm-decas}
    \renewcommand\arraystretch{1.5}
    \begin{ruledtabular}
        \begin{tabular}{cccc}
            Case & A & B & C1 or C2 \\ \hline
            $G_\infty$ & O(3) & O(2) & O(2)$\times$C$_2$ \\
            $(\Gamma_\infty,\alpha_\infty)$ & $(l^\mathcal{P},m)$ & $(|m|,S_m)$ & $(|m|^\mathcal{P},S_m)$ \\
            \multirow{2}{*}{$\tilde{v}_{\Gamma_\infty}$} & \multirow{2}{*}{$v_l$} & $\mathcal{J}^{\frac{1}{2}}(\mathbf{k}')\,\mathcal{J}^{\frac{1}{2}}(\mathbf{k})\sum_{l}v_l\frac{2l+1}{4\pi} \frac{(l-m)!}{(l+m)!}$ & $\mathcal{J}^{\frac{1}{2}}(\mathbf{k}')\,\mathcal{J}^{\frac{1}{2}}(\mathbf{k})\sum_{l^\mathcal{P}}v_l\frac{2l+1}{4\pi} \frac{(l-m)!}{(l+m)!}$ \\ 
                & & $\times P_{lm}(\cos\theta'^*)P_{lm}(\cos\theta^*)$ & $\times P_{lm}(|\cos\theta'^*|)P_{lm}(|\cos\theta^*|)$ \\ 
            $u_{\Gamma_\infty,\alpha_\infty}$ & $Y_{lm}$ & $e^{im\phi^*}$ & $S^\mathcal{P}(m,\theta^*)\,e^{im\phi^*}$ \\
        \end{tabular}
    \end{ruledtabular}
\end{table}

In general cases, we expect the finite-volume potential in \cref{eq:dfvH} can be put in a similar form as \cref{eq:pwe} as follows:
\begin{align}\label{eq:pweinf}
    \tilde{V}(\mathbf{k}',\mathbf{k}) = \sum_{\Gamma_\infty}\tilde{v}_{\Gamma_\infty}(e_n',e_n)\, \sum_{\alpha_\infty}u_{\Gamma_\infty,\alpha_\infty}(\mathbf{n}')\,u^*_{\Gamma_\infty,\alpha_\infty}(\mathbf{n}) \,,
\end{align}
where $\alpha_\infty$ denotes the index of the vector of the irreducible representation $\Gamma_\infty$ of the group $G_\infty$, and the definitions for $u_{\Gamma_\infty,\alpha_\infty}$ and $\tilde{v}_{\Gamma_\infty}$ are summarized in \cref{tab:elm-decas}.
In the table, we also have
\begin{align}\label{eq:elm-sme}
    &S_m = 
    \begin{cases}
        + & m \geq 0 \\
        - & m < 0 \\
    \end{cases}\,, \quad 
    \sum_{l^\mathcal{P}} = \begin{cases}
        \text{ sum over evens} & \mathcal{P}=+ \\
        \text{ sum over odds} & \mathcal{P}=- \\
    \end{cases}\,,
\end{align}
and $S^\mathcal{P}(m,\theta^*)$ is defined via
\begin{align}
    P_{lm}(\cos\theta^*) =
    \begin{cases}
        S^+(m,\theta^*)\, P_{lm}(|\cos\theta^*|) & l \text{ is even} \\
        S^-(m,\theta^*)\, P_{lm}(|\cos\theta^*|) & l \text{ is odd} \\
    \end{cases}\,,
\end{align}
which gives
\begin{align}
    S^+(m,\theta^*) = 
    \begin{cases}
        +1 & \cos\theta^* \geq 0 \\
        (-1)^{m} & \cos \theta^* < 0 \\
    \end{cases}\,, \quad
    S^-(m,\theta^*) = 
    \begin{cases}
        +1 & \cos\theta^* \geq 0 \\
        (-1)^{m+1} & \cos \theta^* < 0 \\
    \end{cases}\,.
\end{align}

In the standard case, \cref{eq:pweinf} becomes the same as \cref{eq:pwe}.
In case B, we now take O(2) as $G_\infty$, and the $\alpha_\infty$-independence of $\tilde{v}_{\Gamma_\infty}$ comes from the invariance of 
\begin{align}
    \frac{(l-m)!}{(l+m)!}\, P_{lm}(\cos\theta'^*)\, P_{lm}(\cos\theta^*)
\end{align}
under $m\to -m$.
In case C, $G_\infty$ is now O(2)$\times$C$_2$, where the $C_2$ symmetry comes from $\mathbf{k}\to \mathbf{P-k}$, i.e., exchanging the momenta of the two equal-mass particles, and is different from the parity symmetry in the usual sense. In the latter case, one is concerned with $\mathbf{k}\to \mathbf{-k}$ and $\mathbf{P}\to \mathbf{-P}$.
It is now $|\cos\theta^*|$ instead of $\cos\theta^*$ independent of $\hat{e}_n$.

In what follows, we will show how to construct the optimal set of basis states maximally reducing the dimension of the finite-volume Hamiltonian in general cases.
The formalism is basically the same as that in Ref.~\cite{Li:2019qvh}, except with a different language introduced above.

%


Now one can introduce 
\begin{align}
    \ket{e_n;\Gamma_\infty,\alpha_\infty} = \sum_{\hat{e}_n}u_{\Gamma_\infty,\alpha_\infty}(\mathbf{n})\,\ket{\mathbf{n}}
\end{align}
to write $V_L$ as
\begin{align}\label{eq:VLinfty}
    V_L=\eta^{-1}\,L^{-3}\sum_{e_n',e_n;\Gamma_\infty} \tilde{v}_{\Gamma_\infty}(e_n',e_n)\,\sum_{\alpha_\infty}\,\ket{e_n';\Gamma_\infty,\alpha_\infty}\bra{e_n;\Gamma_\infty,\alpha_\infty} \,,
\end{align}
and construct the states $\ket{e_n,\Gamma_\infty;\Gamma,f,\alpha}$ via linear combinations 
of $\ket{e_n;\Gamma_\infty,\alpha_\infty}$ as follows:
\begin{align}
    \ket{e_n,\Gamma_\infty;\Gamma,f,\alpha} = \sum_{\alpha_\infty}[C_{\Gamma_\infty}]_{\alpha_\infty;\Gamma,f,\alpha}\,\ket{e_n;\Gamma_\infty,\alpha_\infty} \,,
\end{align}
where $\Gamma$, $f$ and $\alpha$ denote the $\alpha$-th vector of the $f$-th occurrence of the irreducible representation $\Gamma$ reduced from the $\Gamma_\infty$, and the coefficients derived from group theory can be found in \cref{app:SGR}.

One can then define the inner product matrices for these states as
\begin{align}\label{eq:elm-PM}
    [P_{e_n}]_{\Gamma_\infty',\alpha_\infty';\Gamma_\infty,\alpha_\infty}:=\braket{e_n;\Gamma_\infty',\alpha_\infty'|e_n;\Gamma_\infty,\alpha_\infty} \,,
\end{align}
and
\begin{align}
    [P_{e_n;\Gamma,\alpha}]_{\Gamma_\infty',f';\Gamma_\infty,f} &:=\braket{e_n,\Gamma_\infty';\Gamma,f',\alpha|e_n,\Gamma_\infty;\Gamma,f,\alpha} \nonumber\\
    &=\sum_{\alpha_\infty',\alpha_\infty} [C_{\Gamma_\infty'}]^*_{\alpha_\infty';\Gamma,f',\alpha}\,[P_{e_n}]_{\Gamma_\infty',\alpha_\infty';\Gamma_\infty,\alpha_\infty}\,[C_{\Gamma_\infty}]_{\alpha_\infty;\Gamma,f,\alpha} \, .
\end{align}
Using these inner product matrices, one can orthonormalize $\ket{e_n,\Gamma_\infty;\Gamma,f,\alpha}$ to our final basis $\ket{e_n;\Gamma,F,\alpha}$. The Wigner-Eckart theorem only permits the following general form for the $V_L$:
\begin{align}
    V_L=\eta^{-1}\,L^{-3}\sum_{e_n',e_n;\Gamma,F',F}\tilde{v}_{\Gamma,F',F}(e_n',e_n)\sum_\alpha \,\ket{e_n';\Gamma,F',\alpha}\bra{e_n;\Gamma,F,\alpha} \, ,
\end{align}
which, combined with \cref{eq:VLinfty}, leads to
\begin{align}
    &\tilde{v}_{\Gamma,F',F}(e_n',e_n)=\sum_{\Gamma_\infty}\tilde{v}_{\Gamma_\infty}(e_n',e_n)\,[G_{\Gamma_\infty,\Gamma}]_{e_n',F';e_n,F} \,,\nonumber\\
    &[G_{\Gamma_\infty,\Gamma}]_{e_n',F';e_n,F} = \sum_f [M_{\Gamma_\infty;\Gamma,\alpha}]^*_{f;e_n',F'}\,[M_{\Gamma_\infty;\Gamma,\alpha}]_{f;e_n,F} \,, \quad \forall \alpha \,,\nonumber\\
    &[M_{\Gamma_\infty;\Gamma,\alpha}]_{f;e_n,F} = \braket{e_n,\Gamma_\infty;\Gamma,f,\alpha|e_n;\Gamma,F,\alpha} \,.
\end{align}
There are different methods to orthonormalize $\ket{e_n,\Gamma_\infty;\Gamma,f,\alpha}$. We present the result of the eigenmode-based method discussed in Ref.~\cite{Li:2019qvh} as follows:
\begin{align}
    [M_{\Gamma_\infty;\Gamma,\alpha}]_{f;e_n,F} = \sqrt{\lambda^F}\, X^F_{\Gamma_\infty,f} \,,
\end{align}
where $\lambda^F$ and $X^F_{\Gamma_\infty,f}$ are the $F$-th eigenvalue and the $(\Gamma_\infty,f)$-component of the $F$-th eigenvector of the matrix $P_{e_n;\Gamma,\alpha}$ respectively.

\section{Example of isospin-2 $\pi\pi$ scattering}\label{sec:eis}
In this section, following a similar discussion in Sec. IV of Ref.~\cite{Li:2019qvh}, we apply the
formalism developed herein to analyze lattice QCD results for the isospin-2 $\pi\pi$ scattering
system. This time, the moving-frame data is included in the analysis.

As in Ref.~\cite{Li:2019qvh}, the lattice QCD results are from Ref.~\cite{Dudek:2012gj} where an anisotropic action is used.
They quote the anisotropy $\xi={a_s}/{a_t}=3.444(6)$ and the pion mass in lattice units $a_t\, m_\pi = 0.06906(13)$.
The $\pi\pi$-channel is also studied in their other recent works \cite{Wilson:2015dqa,Briceno:2016mjc,Briceno:2017qmb}.
Drawing on the scale setting provided in Ref.~\cite{Briceno:2017qmb}, $a_t^{-1}=5.662\,$GeV, $m_\pi$ is approximately $391\,$MeV.

%

In the analysis performed in Ref.~\cite{Dudek:2012gj}, lattice results above the $4\pi$ threshold
were not included. Since our formalism does not include the four-body contributions, the same cut
is employed.
%

\subsection{The procedures}
As in Ref.~\cite{Li:2019qvh}, we work with dimensionless lattice units. The kinetic energy $h$ is taken as
\begin{align}
  a_t \,h(k) = 2\sqrt{(a_t\, m_\pi)^2+\left( a_t\, k \right)^2}\,,
\end{align}
and when going to the finite-volume system, we have
\begin{align}
  a_t\, k \to a_t\, k_N = \frac{2\pi\, \sqrt{N}}{\xi\,L/a_s}\,,
\end{align}
where $N = \mathbf{n}^2$.
Because the isospin is two, only $s$, $d$, and $g$ waves need to be taken into account, as in Ref.~\cite{Dudek:2012gj}. With the
partial-wave expansion of \cref{eq:pwe}, the partial-wave potentials are taken to be of a simple separable form
\begin{align}
  a_t^{-2}\, v_l(p,k) = \frac{G_l}{(a_t\, m_\pi)^2}\, f_l(p)\, f_l(k) \,,
\end{align}
with
\begin{align}
  f_l(k) = \frac{(d_l\, a_t\, k)^l}{(1+(d_l\, a_t\, k)^2)^{l/2+2}} \,,
\end{align}
with parameters $G_l$ and $d_l$ dimensionless.

The parameters in these potentials were fit to minimize the $\chi^2$ defined by
\begin{equation}\label{eq:chi2}
  \chi^2 = [E_{\text{Sep}}-E_{\text{Lattice}}]^T\, [\mathbb{C}]^{-1}\, [E_{\text{Sep}}-E_{\text{Lattice}}] \, ,
\end{equation}
where $E_{\text{Sep}}-E_{\text{Lattice}}$ denotes the vector of the differences between the
spectrum obtained in the separable potential model and the lattice simulation.  The covariance matrix $\mathbb{C}$ denotes the
covariances in the lattice spectrum of Ref.~\cite{Dudek:2012gj}.

The spectrum was calculated using the method discussed in \cref{sec:pwmems}. 
While a momentum cutoff $N_{\text{cut}}=600$ was used in Ref.~\cite{Li:2019qvh}, we found that  $N_{\text{cut}}=100$ is already enough for the analysis.
Consider a specific level (we choose the highest one of $A_1^+$ in \cref{fig:ERest}) for example, while the lattice level is $0.263773(424)$, the level solved from the Hamiltonian (using the parameters taken from the rest-frame fit in Ref.~\cite{Li:2019qvh}) only shifts around $0.000001$ when $N_{\text{cut}}$ reduces from $600$ to $100$.
Actually, for the lattice size $L\sim 3\,$fm used here, $\frac{2\pi}{L}\sqrt{N_{\text{cut}}}$ are roughly $10$ and $4$\,GeV when $N_{\text{cut}}=600$ and $100$ respectively. So $N_{\text{cut}}=100$ is totally enough here. For studies with larger $L$, however, one needs larger $N_{\text{cut}}$.
What's more, we found $|\chi^2_{N_{\text{cut}}=100}-\chi^2_{N_{\text{cut}}=600}|<0.1$ in the range of parameters of interest.
On the other hand, the values $a_t\, m_\pi = 0.06906(13)$ and $\xi={a_s}/{a_t}=3.444(6)$ may bring appreciable uncertainties to our analysis.
Here we do not consider them, because the analysis based on the L\"{u}scher method implemented in Ref.~\cite{Dudek:2012gj} suggests that they only have a small effect.
The dimensions of the finite-volume Hamiltonian matrices for $N_{\text{cut}}=100$ and 600 are
listed in \cref{tab:dfvh} for each of the irreducible representations considered.  Case B is not
included, as we have $m_1=m_2$ in the current $\pi\pi$ system.
It is notable that the analysis of the moving-frame lattice data corresponds to the C2 case with $\eta=1$.
\begin{table}[tbp]
    \centering
    \caption{The dimensions of the finite-volume Hamiltonian matrices for each of the irreducible
      representations $\Gamma$, for $N_{\text{cut}}=100$ and 600.}\label{tab:dfvh}
    \renewcommand\arraystretch{1.5}
    \begin{ruledtabular}
        \begin{tabular}{cccc}
            Case: $\mathbf{d}$ & $\Gamma$ & $N_{\text{cut}}=100$ & $N_{\text{cut}}=600$ \\ \hline
            A : $(0,0,0)$ & $(A_1^+,A_2^+,E^+,T_1^+,T_2^+)$ & $(129,0,145,75,144)$ & $(923,0,965,488,963)$ \\
            C1: $(0,0,1)$ & $(A_1^+,A_2^+,B_1^+,B_2^+,E^+)$ & $(357,202,271,249,448)$ & $(4357,3004,3354,3254,6222)$ \\
            C1: $(0,1,1)$ & $(A_1^+,A_2^+,B_1^+,B_2^+)$ & $(624,467,465,487)$ & $(8122,6806,6802,6923)$ \\
            C1: $(1,1,1)$ & $(A_1^+,A_2^+,E^+)$ & $(409,239,652)$ & $(5320,3504,8879)$ \\
            C2: $(0,0,1)$ & $(A_1^+,A_2^+,B_1^+,B_2^+,E^+)$ & $(308,173,234,214,448)$ & $(4102,2826,3158,3064,6222)$ \\
            C2: $(0,1,1)$ & $(A_1^+,A_2^+,B_1^+,B_2^+)$ & $(558,420,417,433)$ & $(7772,6516,6518,6625)$ \\
            C2: $(1,1,1)$ & $(A_1^+,A_2^+,E^+)$ & $(354,215,564)$ & $(5035,3360,8381)$ \\
        \end{tabular}
    \end{ruledtabular}
\end{table}

\subsection{The results}
\begin{figure}[ptb]
    \centering
    \includegraphics[width=8cm]{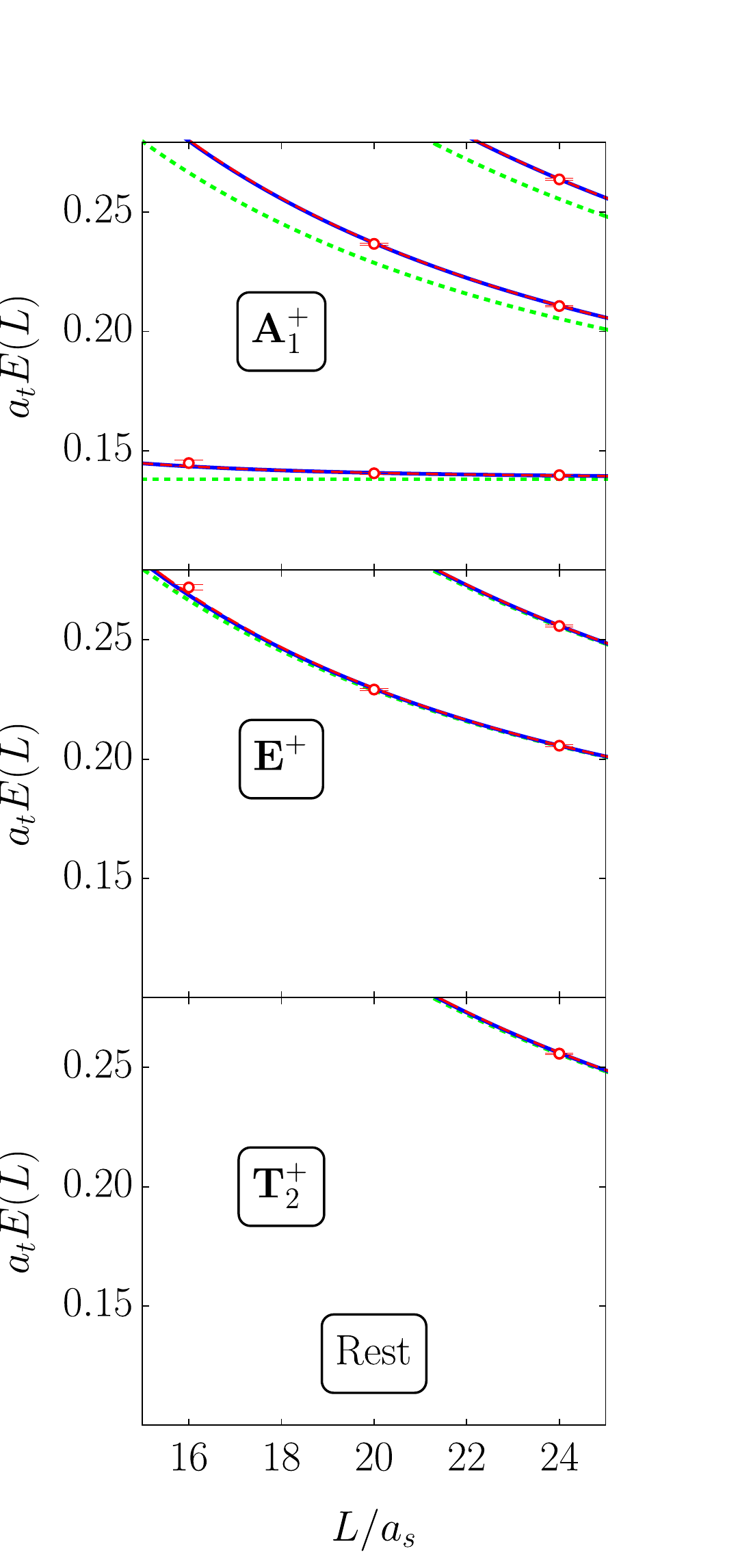}
    \caption{Rest-frame finite-volume spectrum fit of the separable potential model to the lattice
      QCD results of Ref.~\cite{Dudek:2012gj} for isospin-2 $\pi\pi$ scattering. 
          Red dashed (Rest-only fit) and blue solid (Rest-\&-Moving fit) curves illustrate the energies resolved in the separable potential model as the fit parameters of \cref{tab:pmdds} are optimized to fit the lattice QCD results (red points in this figure and blue points in \cref{fig:EMoving}).
          Results of the Rest-only and Rest-\&-Moving fits are almost indistinguishable.
          Green dotted curves illustrate the noninteracting
          pion-pair energies.
      }\label{fig:ERest}
\end{figure}

\begin{figure}[tbp]
    \centering
    \includegraphics[width=16.2cm]{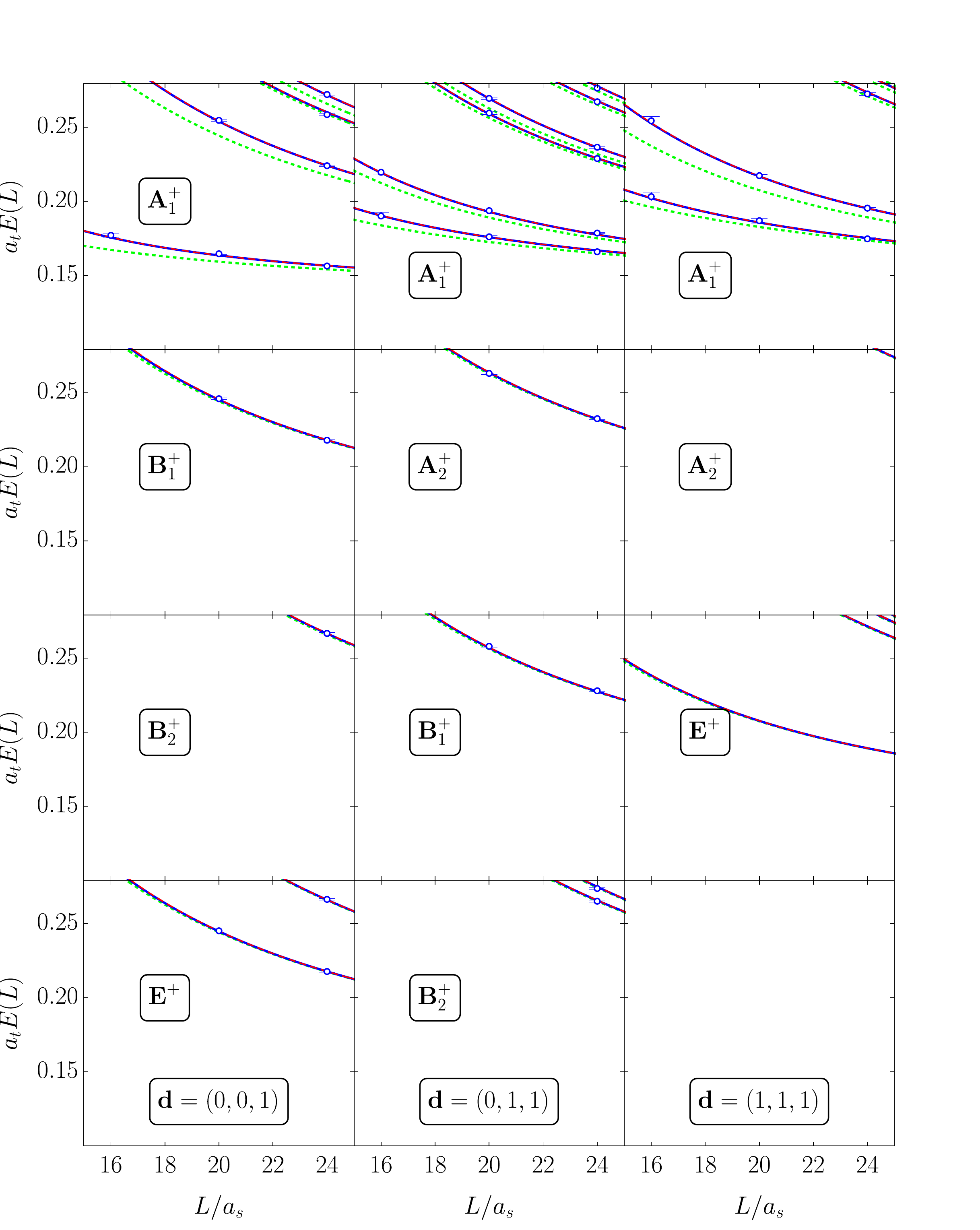}
    \caption{
        As in \cref{fig:ERest} for moving-frame finite-volume spectrum.
      }
\label{fig:EMoving}
\end{figure}

As in Ref.~\cite{Li:2019qvh} we set $d_2=d_4=d_B=4.78$ in the fitting. The results of the fit are shown in \cref{tab:pmdds}, where results of Ref.~\cite{Li:2019qvh} are also included for comparison.
Using those parameters, we predict the $L$-dependent spectrum for both rest and moving frames in \cref{fig:ERest,fig:EMoving}.
\begin{table}[tbp]
    \centering
    \caption{Parameters minimizing \cref{eq:chi2} with rest-frame data only and both rest- and moving-frame data. Both data are from Ref.~\cite{Dudek:2012gj} and the fitting results of rest-frame data are taken from Ref.~\cite{Li:2019qvh}. Covariances for parameters are described by the Hessians listed in \cref{eq:HessianRest,eq:HessianMoving}.}\label{tab:pmdds}
    \renewcommand\arraystretch{1.5}
    \begin{ruledtabular}
        \begin{tabular}{cccccccc}
            &          &\multicolumn{2}{c}{$\ell = 0$} &\multicolumn{2}{c}{$\ell = 2$} &\multicolumn{2}{c}{$\ell = 4$} \\
            Data used & $\chi^2/N_{\text{dof}}$ & $G_0$ & $d_0$ & $G_2$ & $d_2$ & $G_4$ & $d_4$ \\ \hline
            Rest only & 10.5/(11-4) & 67.8(3.4) & 4.57(0.28) & 90.6(28.3) & $d_B$ & 340.(307.) & $d_B$ \\
            Rest \& Moving & 115.9/(49-4) & 67.2(2.3) & 4.59(0.18) & 68.1(16.4) & $d_B$ & 257.(173.) & $d_B$ \\
       \end{tabular}
    \end{ruledtabular}
\end{table}

\begin{figure}[tbp]
    \centering
    \includegraphics[width=7.5cm]{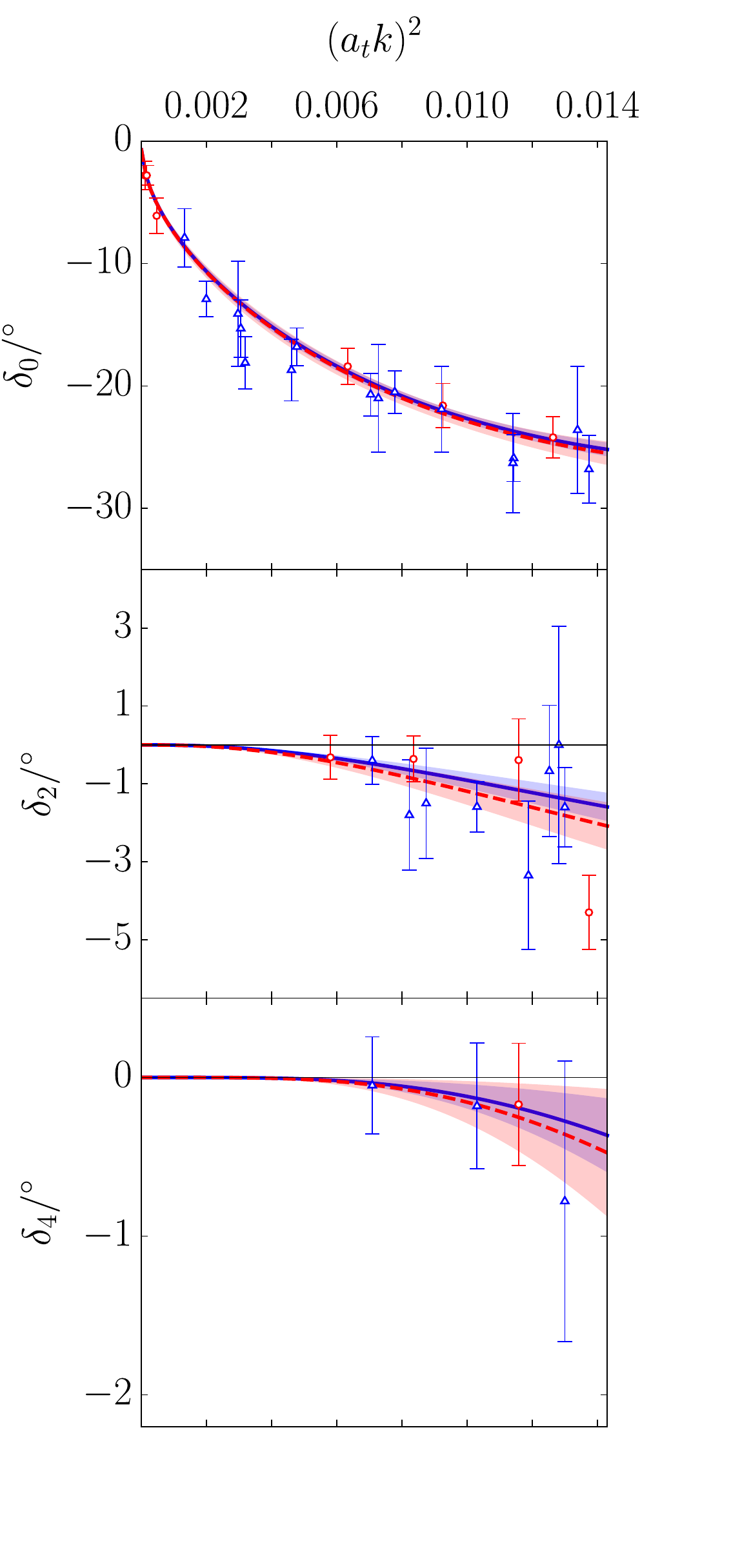}
    \caption{Phase shift curves predicted by the HEFT separable potential model for $s$ (top), $d$
      (middle) and $g$ (bottom) partial waves are compared with points determined via L\"{u}scher's
      method in Ref.~\cite{Dudek:2012gj}.
      Red dashed (blue solid) curves illustrate the central values of the phase shifts obtained
      from rest-frame data only (both rest- and moving-frame data).  The coloured shading describes
      the associated $ 1 \sigma$ uncertainties.
    The red circle (blue triangle) points from Ref.~\cite{Dudek:2012gj} illustrate the phase shifts
    that that can be extracted from the finite-volume spectrum of \cref{fig:ERest,fig:EMoving}
    using L\"{u}scher's method.}\label{fig:delta} 
\end{figure}
\begin{figure}[tbp]
    \centering
    \includegraphics[width=16.2cm]{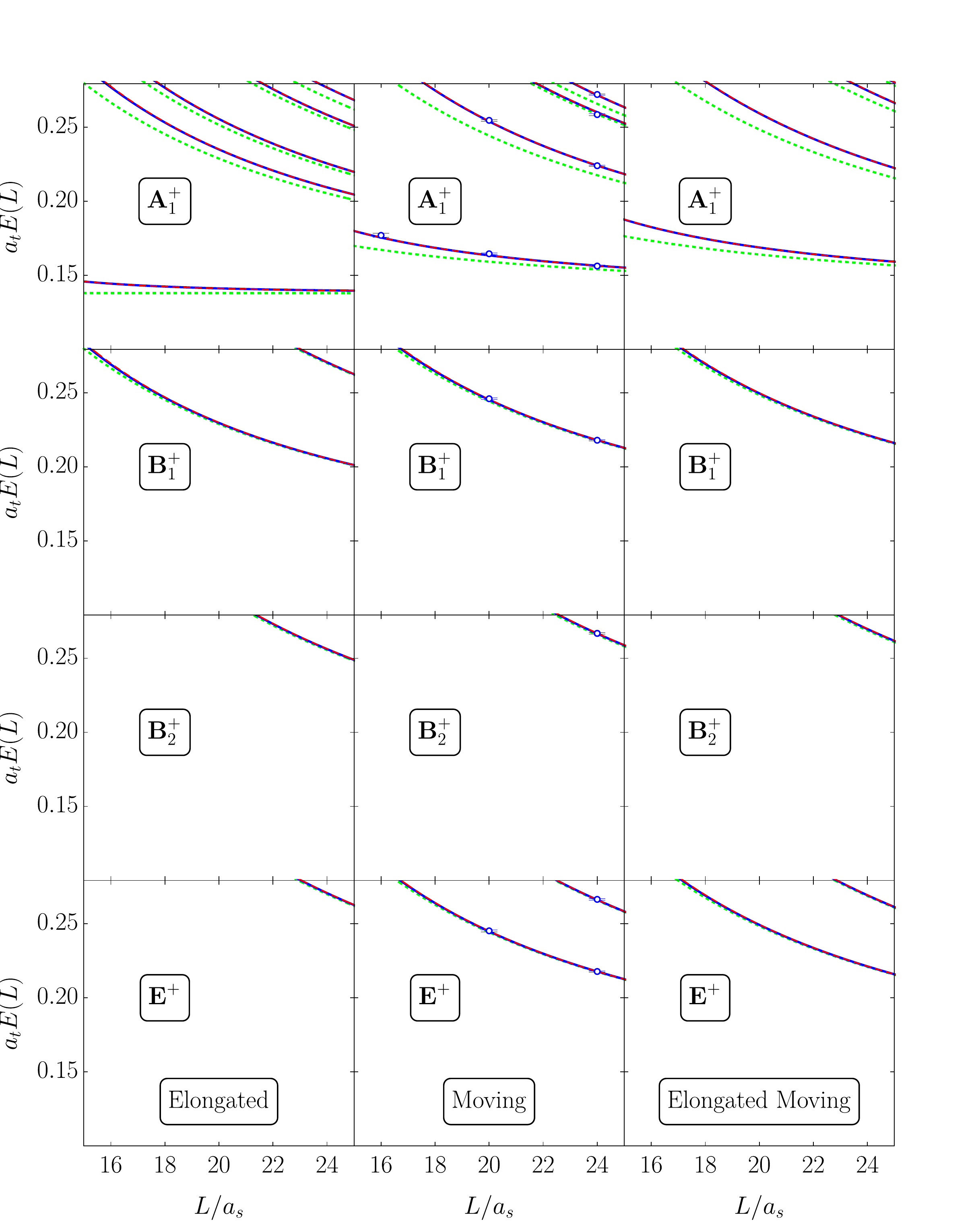}
    \caption{As in \cref{fig:ERest} for elongated, moving and elongated moving systems with
      $\eta\approx0.868$ and $\mathbf{d}=(0,0,1)$.}\label{fig:EEMEM} 
\end{figure}


Our covariance for parameters $\lambda_i$ is defined as $[\mathcal{H}/2]^{-1}$, where $\mathcal{H}$ is the Hessian of $\chi^2$, the matrix of second-order partial derivatives over parameters
\begin{equation}
[\mathcal{H}]_{i,j}=\frac{\partial^2\chi^2}{\partial \lambda_i\, \partial \lambda_j} \, .
\end{equation}
As two $d_l$ are fixed, we only have four parameters and the final covariance returned by MINUIT 2 (ordered as $G_0$, $d_0$, $G_2$, $G_4$) is
\begin{align}\label{eq:HessianMoving}
    [\mathcal{H}/2]^{-1}=
    \left[
    \begin{array}{cccc}
    \underline{5.28} & \underline{0.321} & 21.1 & 47.3 \\
    \underline{0.321} & \underline{0.0323} & 0.839 & 3.78 \\
    21.1 & 0.839 & \underline{269.} & 716. \\
    47.3 & 3.78 & 716. & \underline{2.99\times10^4} \\
    \end{array}
    \right] \,.
\end{align}
For comparison, we also list the covariance obtained in Ref.~\cite{Li:2019qvh} for the fitting of rest-frame data:
\begin{align}\label{eq:HessianRest}
    [\mathcal{H}_{\text{Rest}}/2]^{-1}=
    \left[
    \begin{array}{cccc}
    \underline{11.4} & \underline{0.674} & 58.4 & -197. \\
    \underline{0.674} & \underline{0.0773} & 1.53 & -9.61 \\
    58.4 & 1.53 & \underline{802.} & -1.28\times10^3 \\
    -197. & -9.61 & -1.28\times10^3 & \underline{9.45\times10^4} \\
    \end{array}
    \right] \,.
\end{align}
As the different values of $l$ are decoupling in solving for the phase shifts, the values underlined in \cref{eq:HessianMoving,eq:HessianRest} are used in calculating the errors in the phase shifts.  


In \cref{tab:pmdds}, one may be concerned with the increase of $\chi^2/N_{\text{dof}}$ after including the moving frame data.
However, the fitting based on the L\"{u}scher method implemented in Ref.~\cite{Dudek:2012gj} gives $\chi^2/N_{\text{dof}}=116/(49-3)$ (they did not include the $g$-wave), quite close to ours.
As shown in \cref{fig:delta}, the inclusion of moving frame lattice QCD results induces small
variations in the phase shifts within the $1 \sigma$ uncertainties of the predictions from the
rest-frame data alone.  However, the use of moving-frame data reduces the uncertainties in the HEFT
phase shifts significantly.  Moreover, the constraints provided by several lattice QCD energy
levels on a small number of parameters characterising the spectrum lead to results that are
relatively precise in comparison to the L\"{u}scher method.  Still the results from the two approaches
are generally consistent with only one outlier arising from the L\"{u}scher method.

We also examine the differences in the finite-volume spectra associated with the elongation of the lattice volume versus the nonzero total momentum of the two-particle system.
Our consideration aims to understand how elongation and nonzero total momentum differ in the spectrum.  One may find for the pure elongated and the pure moving systems that $\frac{1}{\eta}$ and $\gamma$ play a similar role. Noting that $\gamma$ depends on the momentum $\mathbf{k}$ and the
total momentum $\mathbf{P}$, to make a comparison we set $\eta=\frac{1}{\gamma}\approx0.868$ with $\gamma$ taking the value on $\mathbf{k}=0$, $\mathbf{d}=(0,0,1)$ and $L/a_s=20$. We predict the $L$-dependent spectrum for the corresponding elongated, moving and elongated moving systems in
\cref{fig:EEMEM}.  These three systems provide quite different spectra.  Thus the consideration of elongated, moving, and elongated-moving systems are useful for generating more data within a certain range of lattice sizes.  

\section{Summary}\label{sec:sum}
In this work, we have extended HEFT to accommodate both an elongated finite volume and systems with
nonzero total momentum. We also consider their combination when the directions of the elongation and
the total momentum are aligned.
To calculate the finite-volume energy levels, we first constructed the elongated-moving Hamiltonian
\cref{eq:mmt,eq:jacob,eq:dfvH,eq:tmp} via the potential parametrized in the rest frame.
The spectrum solved from the Hamiltonian can approximate the real spectrum of the elongated moving
system up to exponentially-suppressed corrections.
The elongation was handled in the usual way, and the moving effects were realized via a momentum
transformation proposed in Ref.~\cite{Li:2021General}.

We then applied the formalism proposed in Ref.~\cite{Li:2019qvh} to disentangle the partial-wave
mixing in the elongated moving Hamiltonian.  This formalism maximally reduces the dimension of the
Hamiltonian matrix.
Different from the rest frame, the elongated moving system has an additional characteristic
direction, which prefers the partial-wave expansion expanded in a specific coordinate system.

Next, an example of isospin-2 $\pi\pi$ scattering was used to demonstrate how this formalism works.
The use of moving-frame data induced small corrections in the phase shifts within the $1\sigma$
bounds of the rest-frame predictions.  However the consideration of moving-frame lattice results
significantly reduced the uncertainty in the phase-shift predictions.
The consistency between analyses from the rest-frame Hamiltonian formalism implemented in
Ref.~\cite{Li:2019qvh} and the L\"{u}scher method implemented in Ref.~\cite{Dudek:2012gj} is
maintained by the moving-frame Hamiltonian formalism implemented here. 

Finally, we examined differences between the effects of elongation and nonzero total momentum. The
spectra obtained from the elongated, moving and elongated moving systems are quite different, and
provide additional avenues for generating more Lattice QCD results within a certain range of
lattice size.
On the current status of lattice simulations, $L$ is roughly in the range $2$-$6\,$fm.

This work has largely accomplished the outlook of Ref.~\cite{Li:2019qvh} for the generalization of the Hamiltonian formalism.
More applications to two-body channels with data from elongated and moving systems are planned.
Furthermore, as mentioned in Ref.~\cite{Li:2019qvh}, the moving-frame formalism
developed here is necessary for a three-body formalism, since two of the three particles can have a
nonvanishing total momentum. In the three-body case, a direct Hamiltonian fit should be formally
simpler than the three-body L\"{u}scher formalism. Of course, one of the remaining challenges is
the significant increase in the dimension of the resultant Hamiltonian matrix.

\section*{Acknowledgements}
The finite-volume energy levels and their covariances from Ref.~\cite{Dudek:2012gj} were provided
by the Hadron Spectrum Collaboration -- no endorsement on their part of the analysis presented in
the current paper should be assumed.
We thank Frank X. Lee, Ross D. Young and James M. Zanotti for comments and discussions.
This work is also supported by the Fundamental Research Funds for the Central Universities.
This research was supported by the Australian Research Council through ARC Discovery Project Grants Nos.\ DP150103101 and DP180100497 (A.W.T.) and DP150103164, DP190102215 and DP210103706 (D.B.L.).

\newpage
\begin{appendix}


\section{Symmetry groups relevant to finite volume}\label{app:SGR}

In the infinite volume, with the symmetry group $G_\infty$ and its irreducible representations (irreps) $\Gamma_\infty$, one can label the quantum state vectors as $\ket{\Gamma_\infty,\alpha_\infty}$, where $\alpha_\infty$ is the index for the vectors in the irrep $\Gamma_\infty$. 
For example, if we consider $G_\infty = SO(3)$, then $\ket{\Gamma_\infty,\alpha_\infty}$ will be $\ket{l,m}$.
In the finite volume, the symmetry group $G_\infty$ reduces into one of its subgroups $G$, and the vectors for an irrep $\Gamma$ are now labelled as $\ket{\Gamma,\alpha}$.

According to the restricted representation, the infinite-volume vectors $\ket{\Gamma_\infty,\alpha_\infty}$ also behave as the vectors belonging to the representations of $G$, hence, they can be combined to obtain the vectors belonging to the irrep $\Gamma$ as follows:
\begin{align}
    \ket{\Gamma,f,\alpha} = \sum_{\alpha_\infty} [C_{\Gamma_\infty}]_{\alpha_{\infty};\Gamma,f,\alpha}\,\ket{\Gamma_{\infty},\alpha_{\infty}} \,,
\end{align}
where $f$ is introduced since a specific $\Gamma$ can be obtained more than once from the reduction of a single $\Gamma_\infty$. $C_{\Gamma_\infty}$ is the unitary coefficient matrix.
The purpose of this section is to provide some frequently used $C_{\Gamma_\infty}$. The results are summarized in \cref{tab:gfv-000Boson,tab:gfv-000Fermion,tab:gfv-001Boson,tab:gfv-001Fermion,tab:gfv-011Boson,tab:gfv-011Fermion,tab:gfv-111Boson,tab:gfv-111Fermion}.


Roughly speaking, bosons and fermions can be classified by the single-valued and double-valued irreps of the group O(3) respectively. The group O(3) is isomorphic to SO(3)$\times$C$_2$ where the C$_2$ is generated by the parity inversion. The relation can be formally written as
\begin{align}
    \begin{bmatrix} 
        &&\\
        &\text{\Large O(3)}&\\
        &&
    \end{bmatrix}
    =
    \begin{bmatrix}
        &&\\
        &\text{\Large SO(3)}&\\
        && 
    \end{bmatrix}
    \text{\Large$\times$}
    \left\{
        \begin{bmatrix} 1&&\\ &1&\\ &&1 \end{bmatrix}
        \,,
        \begin{bmatrix} -1&&\\ &-1&\\ &&-1 \end{bmatrix}
    \right\}
    \,.
\end{align}

In the elongated moving system, one also cares about the group O(2)$\times$C$_2$. It can be formally written as
\begin{align}
    \begin{bmatrix}
        \text{\Large O(2)} & \begin{matrix}\\\\\end{matrix} \\
        \begin{matrix}&&&&\end{matrix} &	1	\\
    \end{bmatrix}
    \text{\Large$\times$}\left\{
    \begin{bmatrix}
        1&&\\
        &1&\\
        &&1
    \end{bmatrix}
    ,
    \begin{bmatrix}
        -1&&\\
        &-1&\\
        &&-1
    \end{bmatrix}
    \right\}
        \,.
\end{align}
The group O(2) is isomorphic to the semidirect product SO(2)$\rtimes$C$_2$ formally written as
\begin{align}
    \begin{bmatrix}
        \text{\Large O(2)} & \begin{matrix}\\\\\end{matrix} \\
        \begin{matrix}&&&&\end{matrix} &	1	\\
    \end{bmatrix}
    =
    \begin{bmatrix}
        \text{\Large SO(2)} & \begin{matrix}\\\\\end{matrix} \\
        \begin{matrix}&&&&\end{matrix} &	1	\\
    \end{bmatrix}
    \text{\Large$\rtimes$}\left\{
    \begin{bmatrix}
        1&&\\
        &1&\\
        &&1
    \end{bmatrix}
    ,
    \begin{bmatrix}
        1&&\\
        &-1&\\
        &&1
    \end{bmatrix}
    \right\}
        \,,
\end{align}
where the C$_2$ is generated by the reflection rather than the parity inversion.

Since the irreps of direct-product groups can be constructed as the tensor products of the irreps of the two original groups, we will focus on the groups without the parity inversion in the following discussions, that are, SO(3) and O(2). Although the irreps of SO(3) and O(2) are both labelled as integers and half-integers, they have different dimensions. For SO(3), $\alpha_\infty$ takes $-\Gamma_\infty,~-\Gamma_\infty+1,~\cdots,\Gamma_\infty$, so the dimension will be $2\Gamma_\infty+1$. For O(2), $\alpha_\infty$ can only take $+$ when $\Gamma_\infty=0$, and can take both $+$ and $-$ in other cases.

In the finite volume, SO(3) will reduce into the octahedral group $O$, and O(2) will reduce into dihedral groups whose orders will depend on the direction of the elongated moving vector $\mathbf{d}$. Groups and their irreps are summarized in \cref{tab:gfv-groups}, where $A$, $B$ and $K$ are labels for one-dimensional irreps, $E$ and $G$ for two-dimensional, $T$ for three-dimensional, and $H$ for four-dimensional.

\begin{table}[tbp]
    \centering
    \caption{Groups and irreps.}\label{tab:gfv-groups}
    \renewcommand\arraystretch{1.5}
    \begin{ruledtabular}
    \begin{tabular}{ccccccc}
        $G_\infty$ & $\Gamma_\infty~(\text{Boson})$ & $\Gamma_\infty~(\text{Fermion})$ & $\mathbf{d}$ & $G$ & $\Gamma~(\text{Boson})$ & $\Gamma~(\text{Fermion})$ \\ \hline
        SO(3) & $0,~1,~2,~\cdots$ & $\frac{1}{2},~\frac{3}{2},~\frac{5}{2},~\cdots$ & $(0,0,0)$ & O & $A_1,~A_2,~E,~T_1,~T_2$ & $G_1,~G_2,~H$  \\
        O(2) & $0,~1,~2,~\cdots$ & $\frac{1}{2},~\frac{3}{2},~\frac{5}{2},~\cdots$ & $(0,0,1)$ & Dih$_{4}$ & $A_1,~A_2,~B_1,~B_2,~E$ & $G_1,~G_2$  \\ 
        & & & $(0,1,1)$ & Dih$_{2}$ & $A_1,~A_2,~B_1,~B_2$ & $G$ \\ 
        & & & $(1,1,1)$ & Dih$_{3}$ & $A_1,~A_2,~E$ & $K_1,~K_2,~G$ \\ 
    \end{tabular}
    \end{ruledtabular}
\end{table}

For SO(3), those $C_{\Gamma_\infty}$ are provided in many papers, {\it e.g.}, Table A.2 (for bosons) and Table A.4 (for fermions) of Ref.~\cite{Bernard:2008ax}. Here we cite their results in \cref{tab:gfv-000Boson,tab:gfv-000Fermion}.

\begin{table}[tbp]
    \centering
    \caption{$C_{\Gamma_\infty}$ for bosonic irreps of SO(3) taken from Table A.2 of Ref.~\cite{Bernard:2008ax}.}\label{tab:gfv-000Boson}
    \renewcommand\arraystretch{1.5}
    \begin{ruledtabular}
        \begin{tabular}{cccc}
            $\Gamma_\infty$ & $\Gamma$ & $\alpha$ & $\sum_{\alpha_\infty} [C_{\Gamma_\infty}]_{\alpha_{\infty};\Gamma,f\equiv1,\alpha}\ket{\Gamma_{\infty},\alpha_{\infty}}$ \\ \hline
            $0$ & $A_1$ & $1$ & $\ket{0,0}$ \\ 
            $1$ & $T_1$ & $1$ & $\frac{1}{\sqrt{2}}(\ket{1,-1}-\ket{1,1})$ \\
                & & $2$ & $\frac{i}{\sqrt{2}}(\ket{1,-1}+\ket{1,1})$ \\
                & & $3$ & $\ket{1,0}$ \\ 
                $2$ & $E$ & $1$ & $\ket{2,0}$ \\
                & & $2$ & $\frac{1}{\sqrt{2}}(\ket{2,-2}+\ket{2,2})$ \\ 
                & $T_2$ & $1$ & $-\frac{1}{\sqrt{2}}(\ket{2,-1}+\ket{2,1})$ \\
                & & $2$ & $\frac{i}{\sqrt{2}}(\ket{2,-1}-\ket{2,1})$ \\
                & & $3$ & $-\frac{1}{\sqrt{2}}(\ket{2,-2}-\ket{2,2})$ \\ 
                $3$ & $A_2$ & $1$ & $\frac{1}{\sqrt{2}}(\ket{3,-2}-\ket{3,2})$ \\ 
                & $T_1$ & $1$ & $\frac{\sqrt{5}}{4}(\ket{3,-3}-\ket{3,3})-\frac{\sqrt{3}}{4}(\ket{3,-1}-\ket{3,1})$ \\
                & & $2$ & $\frac{-i\sqrt{5}}{4}(\ket{3,-3}+\ket{3,3})-\frac{i\sqrt{3}}{4}(\ket{3,-1}+\ket{3,1})$ \\
                & & $3$ & $\ket{3,0}$ \\
                & $T_2$ & $1$ & $-\frac{\sqrt{3}}{4}(\ket{3,-3}-\ket{3,3})-\frac{\sqrt{5}}{4}(\ket{3,-1}-\ket{3,1})$ \\
                & & $2$ & $\frac{-i\sqrt{3}}{4}(\ket{3,-3}+\ket{3,3})+\frac{i\sqrt{5}}{4}(\ket{3,-1}+\ket{3,1})$ \\
                & & $3$ & $\frac{1}{\sqrt{2}}(\ket{3,-2}+\ket{3,2})$ \\ 
                $4$ & ${A}_1$ & $1$ & $\frac{\sqrt{30}}{12}(\ket{4,-4}+\ket{4,4})+\frac{\sqrt{21}}{6}\ket{4,0}$ \\
                & ${E}$ & $1$ & $-\frac{\sqrt{42}}{12}(\ket{4,-4}+\ket{4,4})+\frac{\sqrt{15}}{6}\ket{4,0}$ \\
                & & $2$ & $-\frac{1}{\sqrt{2}}(\ket{4,-2}+\ket{4,2})$ \\ 
                & ${T}_1$ & $1$ & $-\frac{1}{4}(\ket{4,-3}+\ket{4,3})-\frac{\sqrt{7}}{4}(\ket{4,-1}+\ket{4,1})$ \\
                & & $2$ & $\frac{i}{4}(\ket{4,-3}-\ket{4,3})-\frac{i\sqrt{7}}{4}(\ket{4,-1}-\ket{4,1})$ \\
                & & $3$ & $\frac{1}{\sqrt{2}}(\ket{4,-4}-\ket{4,4})$ \\ 
                & ${T}_2$ & $1$ & $\frac{\sqrt{7}}{4}(\ket{4,-3}+\ket{4,3})-\frac{1}{4}(\ket{4,-1}+\ket{4,1})$ \\
                & & $2$ & $\frac{i\sqrt{7}}{4}(\ket{4,-3}-\ket{4,3})+\frac{i}{4}(\ket{4,-1}-\ket{4,1})$ \\
                & & $3$ & $\frac{1}{\sqrt{2}}(\ket{4,-2}-\ket{4,2})$ \\ 
        \end{tabular}
    \end{ruledtabular}
\end{table}

\begin{table}[tbp]
    \centering
    \caption{$C_{\Gamma_\infty}$ for fermionic irreps of SO(3) taken from Table A.4 of Ref.~\cite{Bernard:2008ax}.}\label{tab:gfv-000Fermion}
    \renewcommand\arraystretch{1.5}
    \begin{ruledtabular}
        \begin{tabular}{cccc}
            $\Gamma_\infty$ & $\Gamma$ & $\alpha$ & $\sum_{\alpha_\infty} [C_{\Gamma_\infty}]_{\alpha_{\infty};\Gamma,f\equiv1,\alpha}\ket{\Gamma_{\infty},\alpha_{\infty}}$ \\ \hline
            $\frac{1}{2}$ & $G_1$ & $1$ & $\ket{\frac{1}{2},\frac{1}{2}}$ \\
                &  & $2$ & $\ket{\frac{1}{2},-\frac{1}{2}}$\\
            $\frac{3}{2}$ & $H$ & $1$ & $\ket{\frac{3}{2},\frac{3}{2}}$ \\
                &  & $2$ & $\ket{\frac{3}{2},\frac{1}{2}}$ \\
                &  & $3$ & $\ket{\frac{3}{2},-\frac{1}{2}}$ \\
                &  & $4$ & $\ket{\frac{3}{2},-\frac{3}{2}}$ \\
            $\frac{5}{2}$ & $G_2$ & $1$ & $\frac{\sqrt{30}}{6}\ket{\frac{5}{2},-\frac{3}{2}}-\frac{\sqrt{6}}{6}\ket{\frac{5}{2},\frac{5}{2}}$ \\
                &  & $2$ & $-\frac{\sqrt{6}}{6}\ket{\frac{5}{2},-\frac{5}{2}}+\frac{\sqrt{30}}{6}\ket{\frac{5}{2},\frac{3}{2}}$ \\ 
                & $H$ & $1$ & $-\frac{\sqrt{30}}{6}\ket{\frac{5}{2},-\frac{5}{2}}-\frac{\sqrt{6}}{6}\ket{\frac{5}{2},\frac{3}{2}}$ \\
                &  & $2$ & $\ket{\frac{5}{2},\frac{1}{2}}$ \\
                &  & $3$ & $-\ket{\frac{5}{2},-\frac{1}{2}}$ \\
                &  & $4$ & $\frac{\sqrt{6}}{6}\ket{\frac{5}{2},-\frac{3}{2}}+\frac{\sqrt{30}}{6}\ket{\frac{5}{2},\frac{5}{2}}$ \\
            $\frac{7}{2}$ & $G_1$ & $1$ & $\frac{\sqrt{15}}{6}\ket{\frac{7}{2},-\frac{7}{2}}+\frac{\sqrt{21}}{6}\ket{\frac{7}{2},\frac{1}{2}}$ \\
                &  & $2$ & $-\frac{\sqrt{21}}{6}\ket{\frac{7}{2},-\frac{1}{2}}-\frac{\sqrt{15}}{6}\ket{\frac{7}{2},\frac{7}{2}}$ \\ 
                & $G_2$ & $1$ & $-\frac{1}{2}\ket{\frac{7}{2},-\frac{3}{2}}+\frac{\sqrt{3}}{2}\ket{\frac{7}{2},\frac{5}{2}}$ \\
                &  & $2$ & $-\frac{\sqrt{3}}{2}\ket{\frac{7}{2},-\frac{5}{2}}+\frac{1}{2}\ket{\frac{7}{2},\frac{3}{2}}$ \\
                & $H$ & $1$ & $\frac{1}{2}\ket{\frac{7}{2},-\frac{5}{2}}+\frac{\sqrt{3}}{2}\ket{\frac{7}{2},\frac{3}{2}}$ \\
                &  & $2$ & $\frac{\sqrt{21}}{6}\ket{\frac{7}{2},-\frac{7}{2}}-\frac{\sqrt{15}}{6}\ket{\frac{7}{2},\frac{1}{2}}$ \\
                &  & $3$ & $-\frac{\sqrt{15}}{6}\ket{\frac{7}{2},-\frac{1}{2}}+\frac{\sqrt{21}}{6}\ket{\frac{7}{2},\frac{7}{2}}$ \\
                &  & $4$ & $\frac{\sqrt{3}}{2}\ket{\frac{7}{2},-\frac{3}{2}}+\frac{1}{2}\ket{\frac{7}{2},\frac{5}{2}}$ \\
        \end{tabular}
    \end{ruledtabular}
\end{table}

For O(2), we take the case $\mathbf{d}=(0,0,1)$ for example. The rotations of angles $0,~\frac{\pi}{2},~\pi$ and $\frac{3\pi}{2}$ in the SO(2) and the reflection in C$_2$ will survive, then the resulting group Dih$_4$ can be formally written as C$_4$$\rtimes$C$_2$. So Dih$_4$ can be generated by the $\frac{\pi}{2}$ rotation element $R_\frac{\pi}{2}$ and the reflection element $R$, whose representation matrices can be chosen to be
\begin{align}\label{eq:gfv-remat}
    R_\frac{\pi}{2}^{\Gamma_\infty=0}&=
    \begin{bmatrix}
        1
    \end{bmatrix}\,,\qquad
    R_\frac{\pi}{2}^{\Gamma_\infty\neq 0}=
    \begin{bmatrix}
        e^{i \,\Gamma_\infty \frac{\pi}{2}} & 0 \\
        0 & e^{-i \,\Gamma_\infty \frac{\pi}{2}} 
    \end{bmatrix}\,, \nonumber\\
    R^{\Gamma_\infty=0}&=
    \begin{bmatrix}
        1
    \end{bmatrix}\,,\qquad
    R^{\Gamma_\infty\neq 0}=
    \begin{bmatrix}
        0 & 1 \\
        1 & 0
    \end{bmatrix}\,,
\end{align}
where we represent $\ket{\Gamma_\infty\neq 0,\pm}$ as follows:
\begin{align}
    \ket{\Gamma_\infty\neq 0,-} \sim 
    \begin{bmatrix}
        1 \\ 0
    \end{bmatrix}\,,\qquad
    \ket{\Gamma_\infty\neq 0,+} \sim 
    \begin{bmatrix}
        0 \\ 1
    \end{bmatrix}\,.
\end{align}
Note when $\Gamma_\infty\neq 0$, since $\Gamma_\infty$ and $\Gamma_\infty+4$ share the same representation matrices as indicated in \cref{eq:gfv-remat}, one will have $C_{\Gamma_\infty}=C_{\Gamma_\infty+4}$. In fact, for $\mathbf{d}=(0,1,1)$, it will be $C_{\Gamma_\infty}=C_{\Gamma_\infty+2}$, and for $\mathbf{d}=(1,1,1)$, it will be $C_{\Gamma_\infty}=C_{\Gamma_\infty+3}$. Because those irreps are only one- or two-dimensional, it is easy to find out those $C_{\Gamma_\infty}$.
We also note that since there is still the freedom to choose the representation matrices of the finite-volume irreps, those $C_{\Gamma_\infty}$ are not unique. The results for our choice are summarized in \cref{tab:gfv-001Boson,tab:gfv-001Fermion,tab:gfv-011Boson,tab:gfv-011Fermion,tab:gfv-111Boson,tab:gfv-111Fermion}. 

\begin{table}[tbp]
    \centering
    \caption{$C_{\Gamma_\infty}$ for bosonic irreps of O(2) when $\mathbf{d}=(0,0,1)$, and $C_{\Gamma_\infty}=C_{\Gamma_\infty+4}$ for $\Gamma_\infty\neq 0$.}\label{tab:gfv-001Boson}
    \renewcommand\arraystretch{1.5}
    \begin{ruledtabular}
        \begin{tabular}{cccc}
            $\Gamma_\infty$ & $\Gamma$ & $\alpha$ & $\sum_{\alpha_\infty} [C_{\Gamma_\infty}]_{\alpha_{\infty};\Gamma,f\equiv1,\alpha}\ket{\Gamma_{\infty},\alpha_{\infty}}$ \\ \hline
            $0$ & $A_1$ & $1$ & $\ket{0,+}$ \\ 
            $1$ & $E$ & $1$ & $\ket{1,-}$ \\ 
            & & $2$ & $\ket{1,+}$ \\
            $2$ & $B_1$ & $1$ & $\frac{1}{\sqrt{2}}\left(\ket{2,-}+\ket{2,+}\right) $ \\
            & $B_2$ & $1$ & $\frac{1}{\sqrt{2}}\left(\ket{2,-}-\ket{2,+}\right) $ \\ 
            $3$ & $E$ & $1$ & $\ket{3,+}$ \\
            & & $2$ & $\ket{3,-}$ \\ 
            $4$ & $A_1$ & $1$ & $\frac{1}{\sqrt{2}}\left(\ket{4,-}+\ket{4,+}\right) $ \\
            & $A_2$ & $1$ & $\frac{1}{\sqrt{2}}\left(\ket{4,-}-\ket{4,+}\right) $ \\
        \end{tabular}
    \end{ruledtabular}
\end{table}

\begin{table}[tbp]
    \centering
    \caption{$C_{\Gamma_\infty}$ for fermionic irreps of O(2) when $\mathbf{d}=(0,0,1)$, and $C_{\Gamma_\infty}=C_{\Gamma_\infty+4}$ for $\Gamma_\infty\neq 0$.}\label{tab:gfv-001Fermion}
    \renewcommand\arraystretch{1.5}
    \begin{ruledtabular}
        \begin{tabular}{cccc}
            $\Gamma_\infty$ & $\Gamma$ & $\alpha$ & $\sum_{\alpha_\infty} [C_{\Gamma_\infty}]_{\alpha_{\infty};\Gamma,f\equiv1,\alpha}\ket{\Gamma_{\infty},\alpha_{\infty}}$ \\ \hline
            $\frac{1}{2}$ & $G_1$ & $1$ & $\ket{\frac{1}{2},-}$ \\ 
            & & $2$ & $\ket{\frac{1}{2},+}$ \\
            $\frac{3}{2}$ & $G_2$ & $1$ & $\ket{\frac{3}{2},-}$ \\ 
            & & $2$ & $\ket{\frac{3}{2},+}$ \\ 
            $\frac{5}{2}$ & $G_2$ & $1$ & $\ket{\frac{5}{2},+}$ \\ 
            & & $2$ & $\ket{\frac{5}{2},-}$ \\
            $\frac{7}{2}$ & $G_1$ & $1$ & $\ket{\frac{7}{2},+}$ \\
            & & $2$ & $\ket{\frac{7}{2},-}$ \\
        \end{tabular}
    \end{ruledtabular}
\end{table}

\begin{table}[tbp]
    \centering
    \caption{$C_{\Gamma_\infty}$ for bosonic irreps of O(2) when $\mathbf{d}=(0,1,1)$, and $C_{\Gamma_\infty}=C_{\Gamma_\infty+2}$ for $\Gamma_\infty\neq 0$.}\label{tab:gfv-011Boson}
    \renewcommand\arraystretch{1.5}
    \begin{ruledtabular}
        \begin{tabular}{cccc}
            $\Gamma_\infty$ & $\Gamma$ & $\alpha$ & $\sum_{\alpha_\infty} [C_{\Gamma_\infty}]_{\alpha_{\infty};\Gamma,f\equiv1,\alpha}\ket{\Gamma_{\infty},\alpha_{\infty}}$ \\ \hline
            $0$ & $A_1$ & $1$ & $\ket{0,+}$ \\
            $1$ & $B_1$ & $1$ & $\frac{1}{\sqrt{2}}\left(\ket{1,-}+\ket{1,+}\right)$ \\
            & $B_2$ & $1$ & $\frac{1}{\sqrt{2}}\left(\ket{1,-}-\ket{1,+}\right)$ \\
            $2$ & $A_1$ & $1$ & $\frac{1}{\sqrt{2}}\left(\ket{2,-}+\ket{2,+}\right)$ \\
            & $A_2$ & $1$ & $\frac{1}{\sqrt{2}}\left(\ket{2,-}-\ket{2,+}\right)$ \\
        \end{tabular}
    \end{ruledtabular}
\end{table}

\begin{table}[tbp]
    \centering
    \caption{$C_{\Gamma_\infty}$ for fermionic irreps of O(2) when $\mathbf{d}=(0,1,1)$, and $C_{\Gamma_\infty}=C_{\Gamma_\infty+2}$ for $\Gamma_\infty\neq 0$.}\label{tab:gfv-011Fermion}
    \renewcommand\arraystretch{1.5}
    \begin{ruledtabular}
        \begin{tabular}{cccc}
            $\Gamma_\infty$ & $\Gamma$ & $\alpha$ & $\sum_{\alpha_\infty} [C_{\Gamma_\infty}]_{\alpha_{\infty};\Gamma,f\equiv1,\alpha}\ket{\Gamma_{\infty},\alpha_{\infty}}$ \\ \hline
            $\frac{1}{2}$ & $G$ & $1$ & $\ket{\frac{1}{2},-}$ \\
            & & $2$ & $\ket{\frac{1}{2},+}$ \\
            $\frac{3}{2}$ & $G$ & $1$ & $\ket{\frac{3}{2},+}$ \\ 
            & & $2$ & $\ket{\frac{3}{2},-}$ \\
        \end{tabular}
    \end{ruledtabular}
\end{table}

\begin{table}[tbp]
    \centering
    \caption{$C_{\Gamma_\infty}$ for bosonic irreps of O(2) when $\mathbf{d}=(1,1,1)$, and $C_{\Gamma_\infty}=C_{\Gamma_\infty+3}$ for $\Gamma_\infty\neq 0$.}\label{tab:gfv-111Boson}
    \renewcommand\arraystretch{1.5}
    \begin{ruledtabular}
        \begin{tabular}{cccc}
            $\Gamma_\infty$ & $\Gamma$ & $\alpha$ & $\sum_{\alpha_\infty} [C_{\Gamma_\infty}]_{\alpha_{\infty};\Gamma,f\equiv1,\alpha}\ket{\Gamma_{\infty},\alpha_{\infty}}$ \\ \hline
            $0$ & $A_1$ & $1$ & $\ket{0,+}$ \\
            $1$ & $E$ & $1$ & $\ket{1,-}$ \\ 
            & & $2$ & $\ket{1,+}$ \\ 
            $2$ & $E$ & $1$ & $\ket{2,+}$ \\ 
            &  & $2$ & $\ket{2,-}$ \\ 
            $3$ & $A_1$ & $1$ & $\frac{1}{\sqrt{2}}\left(\ket{3,-}+\ket{3,+}\right)$ \\
            & $A_2$ & $1$ & $\frac{1}{\sqrt{2}}\left(\ket{3,-}-\ket{3,+}\right)$ \\
        \end{tabular}
    \end{ruledtabular}
\end{table}

\begin{table}[tbp]
    \centering
    \caption{$C_{\Gamma_\infty}$ for fermionic irreps of O(2) when $\mathbf{d}=(1,1,1)$, and $C_{\Gamma_\infty}=C_{\Gamma_\infty+3}$ for $\Gamma_\infty\neq 0$.}\label{tab:gfv-111Fermion}
    \renewcommand\arraystretch{1.5}
    \begin{ruledtabular}
        \begin{tabular}{cccc}
            $\Gamma_\infty$ & $\Gamma$ & $\alpha$ & $\sum_{\alpha_\infty} [C_{\Gamma_\infty}]_{\alpha_{\infty};\Gamma,f\equiv1,\alpha}\ket{\Gamma_{\infty},\alpha_{\infty}}$ \\ \hline
            $\frac{1}{2}$ & $G$ & $1$ & $\ket{\frac{1}{2},-}$ \\
            & & $2$ & $\ket{\frac{1}{2},+}$ \\ 
            $\frac{3}{2}$ & $K_1$ & $1$ & $\frac{1}{\sqrt{2}}\left(\ket{\frac{3}{2},-}+\ket{\frac{3}{2},+}\right)$ \\ 
            & $K_2$ & $1$ & $\frac{1}{\sqrt{2}}\left(\ket{\frac{3}{2},-}-\ket{\frac{3}{2},+}\right)$ \\
            $\frac{5}{2}$ & $G$ & $1$ & $\ket{\frac{5}{2},+}$ \\ 
            & & $2$ & $\ket{\frac{5}{2},-}$ \\
        \end{tabular}
    \end{ruledtabular}
\end{table}

\subsection{States in elongated moving system}\label{app:SEMS}
In the elongated moving system, one will deal with the states
\begin{align}
    \ket{e_n;\Gamma_\infty,\alpha_\infty} := \sum_{\hat{e}_n} e^{im\phi}\,\ket{\mathbf{n}} \,,
\end{align}
where $e_n$ denotes $(\mathbf{n}^2,\mathbf{n}\cdot\mathbf{d})$, and $\sum_{\hat{e}_n}$ means summing over all the states with the same $e_n$, and $(\Gamma_\infty,\alpha_\infty)$ will be $(|m|,\text{sign}(m))$ (we define sign$(0)=+$), and the angle $\phi$ depends on the choice of the axes labelled as $(\tilde{\mathbf{x}},\tilde{\mathbf{y}},\tilde{\mathbf{z}})$, which can differ from the finite-volume box's axes $(\mathbf{x},\mathbf{y},\mathbf{z})$.
The purpose of this section is to find out some suitable choices for $(\tilde{\mathbf{x}},\tilde{\mathbf{y}},\tilde{\mathbf{z}})$ so that the representation matrices of the symmetry group are consistent with \cref{eq:gfv-remat} (and its counterparts for other $\mathbf{d}$) and hence the $C_{\Gamma_\infty}$ provided before can be used. 
The results are summarized in \cref{tab:gfv-xyz}.

We take the case $\mathbf{d}=(0,0,1)$ for example as before.
One can first choose $\tilde{\mathbf{z}}$ to be the normalized elongated moving vector $\mathbf{d}/|\mathbf{d}|$, and also chooses the rotation axis of $R_{\phi_0}$ in the SO(2) to be $\tilde{\mathbf{z}}$, then only rotations of angles $0,~\frac{\pi}{2},~\pi \text{ and } \frac{3\pi}{2}$ will always send an integer vector to another integer vector as expected. So one has
\begin{align}
    R_{\phi_0} \,\ket{e_n;\Gamma_\infty,\alpha_\infty} = e^{-im\phi_0} \,\ket{e_n;\Gamma_\infty,\alpha_\infty}
\end{align}
for $\phi_0$ to be $0,~\frac{\pi}{2},~\pi \text{ or } \frac{3\pi}{2}$, which is consistent with \cref{eq:gfv-remat}.
One then chooses the reversion axis of $R$ in the C$_2$ to be $\tilde{\mathbf{x}}$, so $R$ will send $\mathbf{n}$ to $\mathbf{n}-2\left(\mathbf{n}\cdot\tilde{\mathbf{y}}\right)\tilde{\mathbf{y}}$. If $\tilde{\mathbf{y}}$ is chosen to make $2\left(\mathbf{n}\cdot\tilde{\mathbf{y}}\right)\tilde{\mathbf{y}}$ an integer vector for any integer vector $\mathbf{n}$, one will have (for $\Gamma_\infty\neq 0$)
\begin{align}
    R \,\ket{e_n;\Gamma_\infty,\alpha_\infty} = \ket{e_n;\Gamma_\infty,-\alpha_\infty} \,,
\end{align}
which is consistent with \cref{eq:gfv-remat}. Our choices for $(\tilde{\mathbf{x}},\tilde{\mathbf{y}},\tilde{\mathbf{z}})$ are summarized in \cref{tab:gfv-xyz}.

\begin{table}[tbp]
    \centering
    \caption{$(\tilde{\mathbf{x}},\tilde{\mathbf{y}},\tilde{\mathbf{z}})$ for different $\mathbf{d}$}\label{tab:gfv-xyz}
    \renewcommand\arraystretch{1.5}
    \begin{ruledtabular}
        \begin{tabular}{cccc}
            $\mathbf{d}$ & $\tilde{\mathbf{x}}$ & $\tilde{\mathbf{y}}$ & $\tilde{\mathbf{z}}$ \\ \hline
            $(0,0,1)$ & $(1,0,0)$ & $(0,1,0)$ & $(0,0,1)$ \\
            $(0,1,1)$ & $(1,0,0)$ & $\frac{1}{\sqrt{2}}(0,1,-1)$ & $\frac{1}{\sqrt{2}}(0,1,1)$ \\
            $(1,1,1)$ & $\frac{1}{\sqrt{6}}(2,-1,-1)$ & $\frac{1}{\sqrt{2}}(0,1,-1)$ & $\frac{1}{\sqrt{3}}(1,1,1)$ \\ 
        \end{tabular}
    \end{ruledtabular}
\end{table}


\section{Solving for the P matrix}\label{app:spm}

\subsection{Rest-frame P matrix}
With the usual definition of the spherical harmonics,
\begin{align}
    &Y_{lm}(\hat{\mathbf{n}}) = \sqrt{\frac{2l+1}{4\pi}\frac{(l-m)!}{(l+m)!}}\,P_{lm}(\cos\theta)\,e^{im\phi} \,, \nonumber\\
    &P_{lm}(x) = \frac{(-1)^m}{2^l l!} (1-x^2)^{m/2} \frac{d^{l+m}}{dx^{l+m}} (x^2-1)^l 
    \label{eq:APPB1}
\end{align}
and the definition of P matrix given in Ref.~\cite{Li:2019qvh}, the rest-frame P matrix will be 
\begin{align}\label{eq:PlmPlm}
    [P_{\mathbf{n}^2}]_{l',m';l,m} &= 4\pi\sum_{\hat{\mathbf{n}}}Y_{l'm'}^*(\hat{\mathbf{n}})\,Y_{lm}(\hat{\mathbf{n}}) \nonumber\\
    &= c_{l',m';l,m} \sum_{n_z} P_{l'm'}(\cos\theta)\,P_{lm}(\cos\theta) \sum_{n_x,n_y} e^{-i(m'-m)\phi} \,,
\end{align}
where 
\begin{align}
    c_{l',m';l,m} = \sqrt{(2l'+1)\frac{(l'-m')!}{(l'+m')!}}\,\sqrt{(2l+1)\frac{(l-m)!}{(l+m)!}} \,.
\end{align}
When $\mathbf{n}=(0,0,0)$, the direction angles are ill-defined, and we can set $Y_{lm}=\frac{\delta_{l0}}{\sqrt{4\pi}}$.

There are many useful properties of the P matrix listed as follows:
\begin{itemize}
    \item $[P_{\mathbf{n}^2}]_{l',m';l,m}$ is real because of the symmetry under $\phi\to-\phi$.
    \item $[P_{\mathbf{n}^2}]_{l',m';l,m}=0$ when $l'+l$ is odd because of the symmetry under $\mathbf{n}\to-\mathbf{n}$.
    \item $[P_{\mathbf{n}^2}]_{l',m';l,m}=0$ when $m'+m$ is odd because of the symmetry under $\phi\to\phi+\pi$.
    \item $[P_{\mathbf{n}^2}]_{l',m';l,m}=0$ when $|m'-m|=2$ because of the symmetry under $\phi\to\phi+\frac{\pi}{2}$.
\end{itemize}

The result of the summation
\begin{align}\label{eq:pma-ros}
    \sum_{n_x,n_y} e^{-i(m'-m)\phi}
\end{align}
depends on $({\mathbf{n}^2},n_z)$.
One need not calculate it for all $({\mathbf{n}^2},n_z)$ by noting the map $\mathbf{n}\to\pm\mathbf{n}+(0,0,j)$ with  any integer $j$.
In fact, the summation \cref{eq:pma-ros} is related to the P matrix $P^B_{e_n}$ with $\mathbf{d}=(0,0,1)$ discussed in \cref{sec:pems}.
%

\subsection{P matrix of the elongated moving system}\label{sec:pems}
From \cref{eq:elm-PM}, the P matrix for the case B in \cref{tab:elm-decas} is
\begin{align}\label{eq:PMB}
    [P^B_{e_n}]_{|m'|,S_m';|m|,S_m}=\sum_{\hat{e}_n} e^{-i(m'-m)\phi^*} \,,
\end{align}
    where it does not matter how to redefine the ill-defined $\phi^*$ when $\mathbf{n}^2=\mathbf{n_\parallel}^2$, because either $m=m'=0$ then \cref{eq:PMB} is $\phi^*$-independent, or one of $m$ and $m'$ is nonzero then $P_{l'm'}$ or $P_{lm}$ in \cref{tab:elm-decas} and \cref{eq:PlmPlm} vanishes.

To solve for it, it is worth noting that the map $\mathbf{n}\to\mathbf{n}+j\,\mathbf{d}$ with $j\,\mathbf{d}$ any integer vector will tell us
\begin{align}
    P^B_{(\mathbf{n}^2,\mathbf{n}\cdot\mathbf{d})} = P^B_{(\mathbf{n}^2+ 2j\,\mathbf{n}\cdot\mathbf{d}+j^2\mathbf{d}^2,+ \mathbf{n}\cdot\mathbf{d}+j\,\mathbf{d}^2)} \,,
\end{align}
and the map $\mathbf{n}\to-\mathbf{n}+j\,\mathbf{d}$ will tell us
\begin{align}
    P^B_{(\mathbf{n}^2,\mathbf{n}\cdot\mathbf{d})} = (-1)^{m'-m} P^B_{(\mathbf{n}^2- 2j\,\mathbf{n}\cdot\mathbf{d}+j^2\mathbf{d}^2,- \mathbf{n}\cdot\mathbf{d}+j\,\mathbf{d}^2)} \,.
\end{align}

With the coordinate axes $(\tilde{x},\tilde{y},\tilde{z})$ of spherical harmonics taking the values in \cref{tab:gfv-xyz}, there are many useful properties of $P^B_{(\mathbf{n}^2,\mathbf{n}\cdot\mathbf{d})}$ listed as follows:
\begin{itemize}
    \item $P^B_{(\mathbf{n}^2,\mathbf{n}\cdot\mathbf{d})}$ is real because of the symmetry under $\phi\to-\phi$, which holds for all the $\mathbf{d}$ presented in \cref{tab:gfv-xyz}.
    \item $[P^B_{(\mathbf{n}^2,\mathbf{n}\cdot\mathbf{d})}]_{|m'|,S_m';|m|,S_m}=0$ when $m'+m$ is odd because of the symmetry under $\phi\to\phi+\pi$, which holds for $\mathbf{d}=(0,0,1),~(0,1,1)$.
    \item $[P^B_{(\mathbf{n}^2,\mathbf{n}\cdot\mathbf{d})}]_{|m'|,S_m';|m|,S_m}=0$ when $|m'-m|=2$ because of the symmetry under $\phi\to\phi+\frac{\pi}{2}$, which holds only for $\mathbf{d}=(0,0,1)$.
\end{itemize}

For the case C (C1 or C2), the map $\mathbf{n}\to\mathbf{d}_\gamma-\mathbf{n}$ will tell us
\begin{align}
    [P^C_{e_n}]_{|m'|^\pm,S_m';|m|^\mp,S_m} = 0 \,,
\end{align}
and 
\begin{align}
    [P^C_{(\mathbf{n}^2,\{\mathbf{n}\cdot\mathbf{d},(\mathbf{d}_\gamma-\mathbf{n})\cdot\mathbf{d}\})}]_{|m'|^\pm,S_m';|m|^\pm,S_m} =
    \begin{cases}
        2[P^B_{(\mathbf{n}^2,\mathbf{n}\cdot\mathbf{d})}]_{|m'|,S_m';|m|,S_m} & \mathbf{n}\cdot\mathbf{d}>\frac{\mathbf{d}_\gamma^2}{2} \\
        [P^B_{(\mathbf{n}^2,\mathbf{n}\cdot\mathbf{d})}]_{|m'|,S_m';|m|,S_m} & \mathbf{n}\cdot\mathbf{d}=\frac{\mathbf{d}_\gamma^2}{2} \\
        2(-1)^{m'-m}[P^B_{(\mathbf{n}^2,\mathbf{n}\cdot\mathbf{d})}]_{|m'|,S_m';|m|,S_m} & \mathbf{n}\cdot\mathbf{d}<\frac{\mathbf{d}_\gamma^2}{2} \\
    \end{cases}\,,
\end{align}
where we have used that $\cos\theta^*$ has the same sign with $\mathbf{n}\cdot\mathbf{d}-\frac{\mathbf{d}_\gamma^2}{2}$ in case C. We also emphasize that $P^{C1}\neq P^{C2}$ since $\mathbf{d}_\gamma=0$ in case C1 while $\mathbf{d}_\gamma=\mathbf{d}$ in case C2.

\end{appendix}

\newpage
\bibliography{refs}

\begin{thebibliography}{77}%
\makeatletter
\providecommand \@ifxundefined [1]{%
 \@ifx{#1\undefined}
}%
\providecommand \@ifnum [1]{%
 \ifnum #1\expandafter \@firstoftwo
 \else \expandafter \@secondoftwo
 \fi
}%
\providecommand \@ifx [1]{%
 \ifx #1\expandafter \@firstoftwo
 \else \expandafter \@secondoftwo
 \fi
}%
\providecommand \natexlab [1]{#1}%
\providecommand \enquote  [1]{``#1''}%
\providecommand \bibnamefont  [1]{#1}%
\providecommand \bibfnamefont [1]{#1}%
\providecommand \citenamefont [1]{#1}%
\providecommand \href@noop [0]{\@secondoftwo}%
\providecommand \href [0]{\begingroup \@sanitize@url \@href}%
\providecommand \@href[1]{\@@startlink{#1}\@@href}%
\providecommand \@@href[1]{\endgroup#1\@@endlink}%
\providecommand \@sanitize@url [0]{\catcode `\\12\catcode `\$12\catcode
  `\&12\catcode `\#12\catcode `\^12\catcode `\_12\catcode `\%12\relax}%
\providecommand \@@startlink[1]{}%
\providecommand \@@endlink[0]{}%
\providecommand \url  [0]{\begingroup\@sanitize@url \@url }%
\providecommand \@url [1]{\endgroup\@href {#1}{\urlprefix }}%
\providecommand \urlprefix  [0]{URL }%
\providecommand \Eprint [0]{\href }%
\providecommand \doibase [0]{http://dx.doi.org/}%
\providecommand \selectlanguage [0]{\@gobble}%
\providecommand \bibinfo  [0]{\@secondoftwo}%
\providecommand \bibfield  [0]{\@secondoftwo}%
\providecommand \translation [1]{[#1]}%
\providecommand \BibitemOpen [0]{}%
\providecommand \bibitemStop [0]{}%
\providecommand \bibitemNoStop [0]{.\EOS\space}%
\providecommand \EOS [0]{\spacefactor3000\relax}%
\providecommand \BibitemShut  [1]{\csname bibitem#1\endcsname}%
\let\auto@bib@innerbib\@empty
\bibitem [{\citenamefont {L{\"u}scher}(1986{\natexlab{a}})}]{Luscher:1985dn}%
  \BibitemOpen
  \bibfield  {author} {\bibinfo {author} {\bibfnamefont {M.}~\bibnamefont
  {L{\"u}scher}},\ }\href {\doibase 10.1007/BF01211589} {\bibfield  {journal}
  {\bibinfo  {journal} {Commun.Math. Phys.}\ }\textbf {\bibinfo {volume}
  {104}},\ \bibinfo {pages} {177} (\bibinfo {year}
  {1986}{\natexlab{a}})}\BibitemShut {NoStop}%
\bibitem [{\citenamefont {L{\"u}scher}(1986{\natexlab{b}})}]{Luscher:1986pf}%
  \BibitemOpen
  \bibfield  {author} {\bibinfo {author} {\bibfnamefont {M.}~\bibnamefont
  {L{\"u}scher}},\ }\href {\doibase 10.1007/BF01211097} {\bibfield  {journal}
  {\bibinfo  {journal} {Commun.Math. Phys.}\ }\textbf {\bibinfo {volume}
  {105}},\ \bibinfo {pages} {153} (\bibinfo {year}
  {1986}{\natexlab{b}})}\BibitemShut {NoStop}%
\bibitem [{\citenamefont {L{\"u}scher}(1991)}]{Luscher:1990ux}%
  \BibitemOpen
  \bibfield  {author} {\bibinfo {author} {\bibfnamefont {M.}~\bibnamefont
  {L{\"u}scher}},\ }\href {\doibase 10.1016/0550-3213(91)90366-6} {\bibfield
  {journal} {\bibinfo  {journal} {Nuclear Physics B}\ }\textbf {\bibinfo
  {volume} {354}},\ \bibinfo {pages} {531} (\bibinfo {year}
  {1991})}\BibitemShut {NoStop}%
\bibitem [{\citenamefont {Hall}\ \emph {et~al.}(2013)\citenamefont {Hall},
  \citenamefont {Hsu}, \citenamefont {Leinweber}, \citenamefont {Thomas},\ and\
  \citenamefont {Young}}]{Hall:2013qba}%
  \BibitemOpen
  \bibfield  {author} {\bibinfo {author} {\bibfnamefont {J.~M.~M.}\
  \bibnamefont {Hall}}, \bibinfo {author} {\bibfnamefont {A.~C.-P.}\
  \bibnamefont {Hsu}}, \bibinfo {author} {\bibfnamefont {D.~B.}\ \bibnamefont
  {Leinweber}}, \bibinfo {author} {\bibfnamefont {A.~W.}\ \bibnamefont
  {Thomas}}, \ and\ \bibinfo {author} {\bibfnamefont {R.~D.}\ \bibnamefont
  {Young}},\ }\href {\doibase 10.1103/PhysRevD.87.094510} {\bibfield  {journal}
  {\bibinfo  {journal} {Phys. Rev. D}\ }\textbf {\bibinfo {volume} {87}},\
  \bibinfo {pages} {094510} (\bibinfo {year} {2013})},\ \Eprint
  {http://arxiv.org/abs/1303.4157} {arXiv:1303.4157 [hep-lat]} \BibitemShut
  {NoStop}%
\bibitem [{\citenamefont {Hall}\ \emph {et~al.}(2015)\citenamefont {Hall},
  \citenamefont {Kamleh}, \citenamefont {Leinweber}, \citenamefont {Menadue},
  \citenamefont {Owen}, \citenamefont {Thomas},\ and\ \citenamefont
  {Young}}]{Hall:2014uca}%
  \BibitemOpen
  \bibfield  {author} {\bibinfo {author} {\bibfnamefont {J.~M.~M.}\
  \bibnamefont {Hall}}, \bibinfo {author} {\bibfnamefont {W.}~\bibnamefont
  {Kamleh}}, \bibinfo {author} {\bibfnamefont {D.~B.}\ \bibnamefont
  {Leinweber}}, \bibinfo {author} {\bibfnamefont {B.~J.}\ \bibnamefont
  {Menadue}}, \bibinfo {author} {\bibfnamefont {B.~J.}\ \bibnamefont {Owen}},
  \bibinfo {author} {\bibfnamefont {A.~W.}\ \bibnamefont {Thomas}}, \ and\
  \bibinfo {author} {\bibfnamefont {R.~D.}\ \bibnamefont {Young}},\ }\href
  {\doibase 10.1103/PhysRevLett.114.132002} {\bibfield  {journal} {\bibinfo
  {journal} {Phys. Rev. Lett.}\ }\textbf {\bibinfo {volume} {114}},\ \bibinfo
  {pages} {132002} (\bibinfo {year} {2015})}\BibitemShut {NoStop}%
\bibitem [{\citenamefont {Wu}\ \emph {et~al.}(2014)\citenamefont {Wu},
  \citenamefont {Lee}, \citenamefont {Thomas},\ and\ \citenamefont
  {Young}}]{Wu:2014vma}%
  \BibitemOpen
  \bibfield  {author} {\bibinfo {author} {\bibfnamefont {J.-J.}\ \bibnamefont
  {Wu}}, \bibinfo {author} {\bibfnamefont {T.-S.~H.}\ \bibnamefont {Lee}},
  \bibinfo {author} {\bibfnamefont {A.~W.}\ \bibnamefont {Thomas}}, \ and\
  \bibinfo {author} {\bibfnamefont {R.~D.}\ \bibnamefont {Young}},\ }\href
  {\doibase 10.1103/PhysRevC.90.055206} {\bibfield  {journal} {\bibinfo
  {journal} {Phys. Rev. C}\ }\textbf {\bibinfo {volume} {90}},\ \bibinfo
  {pages} {055206} (\bibinfo {year} {2014})},\ \Eprint
  {http://arxiv.org/abs/1402.4868} {arXiv:1402.4868 [hep-lat]} \BibitemShut
  {NoStop}%
\bibitem [{\citenamefont {Liu}\ \emph {et~al.}(2016)\citenamefont {Liu},
  \citenamefont {Kamleh}, \citenamefont {Leinweber}, \citenamefont {Stokes},
  \citenamefont {Thomas},\ and\ \citenamefont {Wu}}]{Liu:2015ktc}%
  \BibitemOpen
  \bibfield  {author} {\bibinfo {author} {\bibfnamefont {Z.-W.}\ \bibnamefont
  {Liu}}, \bibinfo {author} {\bibfnamefont {W.}~\bibnamefont {Kamleh}},
  \bibinfo {author} {\bibfnamefont {D.~B.}\ \bibnamefont {Leinweber}}, \bibinfo
  {author} {\bibfnamefont {F.~M.}\ \bibnamefont {Stokes}}, \bibinfo {author}
  {\bibfnamefont {A.~W.}\ \bibnamefont {Thomas}}, \ and\ \bibinfo {author}
  {\bibfnamefont {J.-J.}\ \bibnamefont {Wu}},\ }\href {\doibase
  10.1103/PhysRevLett.116.082004} {\bibfield  {journal} {\bibinfo  {journal}
  {Phys. Rev. Lett.}\ }\textbf {\bibinfo {volume} {116}},\ \bibinfo {pages}
  {082004} (\bibinfo {year} {2016})},\ \Eprint
  {http://arxiv.org/abs/1512.00140} {arXiv:1512.00140 [hep-lat]} \BibitemShut
  {NoStop}%
\bibitem [{\citenamefont {Liu}\ \emph {et~al.}(2017{\natexlab{a}})\citenamefont
  {Liu}, \citenamefont {Kamleh}, \citenamefont {Leinweber}, \citenamefont
  {Stokes}, \citenamefont {Thomas},\ and\ \citenamefont {Wu}}]{Liu:2016uzk}%
  \BibitemOpen
  \bibfield  {author} {\bibinfo {author} {\bibfnamefont {Z.-W.}\ \bibnamefont
  {Liu}}, \bibinfo {author} {\bibfnamefont {W.}~\bibnamefont {Kamleh}},
  \bibinfo {author} {\bibfnamefont {D.~B.}\ \bibnamefont {Leinweber}}, \bibinfo
  {author} {\bibfnamefont {F.~M.}\ \bibnamefont {Stokes}}, \bibinfo {author}
  {\bibfnamefont {A.~W.}\ \bibnamefont {Thomas}}, \ and\ \bibinfo {author}
  {\bibfnamefont {J.-J.}\ \bibnamefont {Wu}},\ }\href {\doibase
  10.1103/PhysRevD.95.034034} {\bibfield  {journal} {\bibinfo  {journal} {Phys.
  Rev. D}\ }\textbf {\bibinfo {volume} {95}},\ \bibinfo {pages} {034034}
  (\bibinfo {year} {2017}{\natexlab{a}})}\BibitemShut {NoStop}%
\bibitem [{\citenamefont {Liu}\ \emph {et~al.}(2017{\natexlab{b}})\citenamefont
  {Liu}, \citenamefont {Hall}, \citenamefont {Leinweber}, \citenamefont
  {Thomas},\ and\ \citenamefont {Wu}}]{Liu:2016wxq}%
  \BibitemOpen
  \bibfield  {author} {\bibinfo {author} {\bibfnamefont {Z.-W.}\ \bibnamefont
  {Liu}}, \bibinfo {author} {\bibfnamefont {J.~M.~M.}\ \bibnamefont {Hall}},
  \bibinfo {author} {\bibfnamefont {D.~B.}\ \bibnamefont {Leinweber}}, \bibinfo
  {author} {\bibfnamefont {A.~W.}\ \bibnamefont {Thomas}}, \ and\ \bibinfo
  {author} {\bibfnamefont {J.-J.}\ \bibnamefont {Wu}},\ }\href {\doibase
  10.1103/PhysRevD.95.014506} {\bibfield  {journal} {\bibinfo  {journal} {Phys.
  Rev. D}\ }\textbf {\bibinfo {volume} {95}},\ \bibinfo {pages} {014506}
  (\bibinfo {year} {2017}{\natexlab{b}})}\BibitemShut {NoStop}%
\bibitem [{\citenamefont {Wu}\ \emph {et~al.}(2017)\citenamefont {Wu},
  \citenamefont {Kamano}, \citenamefont {Lee}, \citenamefont {Leinweber},\ and\
  \citenamefont {Thomas}}]{Wu:2016ixr}%
  \BibitemOpen
  \bibfield  {author} {\bibinfo {author} {\bibfnamefont {J.-J.}\ \bibnamefont
  {Wu}}, \bibinfo {author} {\bibfnamefont {H.}~\bibnamefont {Kamano}}, \bibinfo
  {author} {\bibfnamefont {T.-S.~H.}\ \bibnamefont {Lee}}, \bibinfo {author}
  {\bibfnamefont {D.~B.}\ \bibnamefont {Leinweber}}, \ and\ \bibinfo {author}
  {\bibfnamefont {A.~W.}\ \bibnamefont {Thomas}},\ }\href {\doibase
  10.1103/PhysRevD.95.114507} {\bibfield  {journal} {\bibinfo  {journal} {Phys.
  Rev. D}\ }\textbf {\bibinfo {volume} {95}},\ \bibinfo {pages} {114507}
  (\bibinfo {year} {2017})},\ \Eprint {http://arxiv.org/abs/1611.05970}
  {arXiv:1611.05970 [hep-lat]} \BibitemShut {NoStop}%
\bibitem [{\citenamefont {Wu}\ \emph {et~al.}(2018)\citenamefont {Wu},
  \citenamefont {Leinweber}, \citenamefont {Liu},\ and\ \citenamefont
  {Thomas}}]{Wu:2017qve}%
  \BibitemOpen
  \bibfield  {author} {\bibinfo {author} {\bibfnamefont {J.-J.}\ \bibnamefont
  {Wu}}, \bibinfo {author} {\bibfnamefont {D.~B.}\ \bibnamefont {Leinweber}},
  \bibinfo {author} {\bibfnamefont {Z.-W.}\ \bibnamefont {Liu}}, \ and\
  \bibinfo {author} {\bibfnamefont {A.~W.}\ \bibnamefont {Thomas}},\ }\href
  {\doibase 10.1103/PhysRevD.97.094509} {\bibfield  {journal} {\bibinfo
  {journal} {Phys. Rev. D}\ }\textbf {\bibinfo {volume} {97}},\ \bibinfo
  {pages} {094509} (\bibinfo {year} {2018})},\ \Eprint
  {http://arxiv.org/abs/1703.10715} {arXiv:1703.10715 [nucl-th]} \BibitemShut
  {NoStop}%
\bibitem [{\citenamefont {Li}\ \emph {et~al.}(2020)\citenamefont {Li},
  \citenamefont {Wu}, \citenamefont {Abell}, \citenamefont {Leinweber},\ and\
  \citenamefont {Thomas}}]{Li:2019qvh}%
  \BibitemOpen
  \bibfield  {author} {\bibinfo {author} {\bibfnamefont {Y.}~\bibnamefont
  {Li}}, \bibinfo {author} {\bibfnamefont {J.-J.}\ \bibnamefont {Wu}}, \bibinfo
  {author} {\bibfnamefont {C.~D.}\ \bibnamefont {Abell}}, \bibinfo {author}
  {\bibfnamefont {D.~B.}\ \bibnamefont {Leinweber}}, \ and\ \bibinfo {author}
  {\bibfnamefont {A.~W.}\ \bibnamefont {Thomas}},\ }\href {\doibase
  10.1103/PhysRevD.101.114501} {\bibfield  {journal} {\bibinfo  {journal}
  {Phys. Rev. D}\ }\textbf {\bibinfo {volume} {101}},\ \bibinfo {pages}
  {114501} (\bibinfo {year} {2020})}\BibitemShut {NoStop}%
\bibitem [{\citenamefont {Feng}\ \emph {et~al.}(2004)\citenamefont {Feng},
  \citenamefont {Li},\ and\ \citenamefont {Liu}}]{Feng:2004ua}%
  \BibitemOpen
  \bibfield  {author} {\bibinfo {author} {\bibfnamefont {X.}~\bibnamefont
  {Feng}}, \bibinfo {author} {\bibfnamefont {X.}~\bibnamefont {Li}}, \ and\
  \bibinfo {author} {\bibfnamefont {C.}~\bibnamefont {Liu}},\ }\href {\doibase
  10.1103/PhysRevD.70.014505} {\bibfield  {journal} {\bibinfo  {journal} {Phys.
  Rev. D}\ }\textbf {\bibinfo {volume} {70}},\ \bibinfo {pages} {014505}
  (\bibinfo {year} {2004})},\ \Eprint {http://arxiv.org/abs/hep-lat/0404001}
  {arXiv:hep-lat/0404001} \BibitemShut {NoStop}%
\bibitem [{\citenamefont {Li}\ and\ \citenamefont {Liu}(2004)}]{Li:2003jn}%
  \BibitemOpen
  \bibfield  {author} {\bibinfo {author} {\bibfnamefont {X.}~\bibnamefont
  {Li}}\ and\ \bibinfo {author} {\bibfnamefont {C.}~\bibnamefont {Liu}},\
  }\href {\doibase 10.1016/j.physletb.2004.02.068} {\bibfield  {journal}
  {\bibinfo  {journal} {Phys. Lett.}\ }\textbf {\bibinfo {volume} {B587}},\
  \bibinfo {pages} {100} (\bibinfo {year} {2004})},\ \Eprint
  {http://arxiv.org/abs/hep-lat/0311035} {arXiv:hep-lat/0311035} \BibitemShut
  {NoStop}%
\bibitem [{\citenamefont {Li}\ \emph {et~al.}(2007)\citenamefont {Li} \emph
  {et~al.}}]{Li:2007ey}%
  \BibitemOpen
  \bibfield  {author} {\bibinfo {author} {\bibfnamefont {X.}~\bibnamefont {Li}}
  \emph {et~al.} (\bibinfo {collaboration} {CLQCD}),\ }\href {\doibase
  10.1088/1126-6708/2007/06/053} {\bibfield  {journal} {\bibinfo  {journal}
  {JHEP}\ }\textbf {\bibinfo {volume} {06}},\ \bibinfo {pages} {053} (\bibinfo
  {year} {2007})},\ \Eprint {http://arxiv.org/abs/hep-lat/0703015}
  {arXiv:hep-lat/0703015} \BibitemShut {NoStop}%
\bibitem [{\citenamefont {Lee}\ and\ \citenamefont
  {Alexandru}(2017)}]{Lee:2017igf}%
  \BibitemOpen
  \bibfield  {author} {\bibinfo {author} {\bibfnamefont {F.~X.}\ \bibnamefont
  {Lee}}\ and\ \bibinfo {author} {\bibfnamefont {A.}~\bibnamefont
  {Alexandru}},\ }\href {\doibase 10.1103/PhysRevD.96.054508} {\bibfield
  {journal} {\bibinfo  {journal} {Phys. Rev. D}\ }\textbf {\bibinfo {volume}
  {96}},\ \bibinfo {pages} {054508} (\bibinfo {year} {2017})},\ \Eprint
  {http://arxiv.org/abs/1706.00262} {arXiv:1706.00262 [hep-lat]} \BibitemShut
  {NoStop}%
\bibitem [{\citenamefont {Li}\ \emph {et~al.}(2018)\citenamefont {Li},
  \citenamefont {Wu},\ and\ \citenamefont {Liu}}]{Li:2018tzx}%
  \BibitemOpen
  \bibfield  {author} {\bibinfo {author} {\bibfnamefont {N.}~\bibnamefont
  {Li}}, \bibinfo {author} {\bibfnamefont {Y.-J.}\ \bibnamefont {Wu}}, \ and\
  \bibinfo {author} {\bibfnamefont {Z.-W.}\ \bibnamefont {Liu}},\ }\href
  {\doibase 10.1103/PhysRevD.97.014509} {\bibfield  {journal} {\bibinfo
  {journal} {Phys. Rev. D}\ }\textbf {\bibinfo {volume} {97}},\ \bibinfo
  {pages} {014509} (\bibinfo {year} {2018})}\BibitemShut {NoStop}%
\bibitem [{\citenamefont {Meng}\ \emph {et~al.}(2009)\citenamefont {Meng} \emph
  {et~al.}}]{Meng:2009qt}%
  \BibitemOpen
  \bibfield  {author} {\bibinfo {author} {\bibfnamefont {G.-Z.}\ \bibnamefont
  {Meng}} \emph {et~al.} (\bibinfo {collaboration} {CLQCD}),\ }\href {\doibase
  10.1103/PhysRevD.80.034503} {\bibfield  {journal} {\bibinfo  {journal} {Phys.
  Rev. D}\ }\textbf {\bibinfo {volume} {80}},\ \bibinfo {pages} {034503}
  (\bibinfo {year} {2009})},\ \Eprint {http://arxiv.org/abs/0905.0752}
  {arXiv:0905.0752 [hep-lat]} \BibitemShut {NoStop}%
\bibitem [{\citenamefont {Pelissier}\ and\ \citenamefont
  {Alexandru}(2013)}]{Pelissier:2012pi}%
  \BibitemOpen
  \bibfield  {author} {\bibinfo {author} {\bibfnamefont {C.}~\bibnamefont
  {Pelissier}}\ and\ \bibinfo {author} {\bibfnamefont {A.}~\bibnamefont
  {Alexandru}},\ }\href {\doibase 10.1103/PhysRevD.87.014503} {\bibfield
  {journal} {\bibinfo  {journal} {Phys. Rev. D}\ }\textbf {\bibinfo {volume}
  {87}},\ \bibinfo {pages} {014503} (\bibinfo {year} {2013})}\BibitemShut
  {NoStop}%
\bibitem [{\citenamefont {Guo}\ \emph {et~al.}(2016)\citenamefont {Guo},
  \citenamefont {Alexandru}, \citenamefont {Molina},\ and\ \citenamefont
  {D{\"o}ring}}]{Guo:2016zos}%
  \BibitemOpen
  \bibfield  {author} {\bibinfo {author} {\bibfnamefont {D.}~\bibnamefont
  {Guo}}, \bibinfo {author} {\bibfnamefont {A.}~\bibnamefont {Alexandru}},
  \bibinfo {author} {\bibfnamefont {R.}~\bibnamefont {Molina}}, \ and\ \bibinfo
  {author} {\bibfnamefont {M.}~\bibnamefont {D{\"o}ring}},\ }\href {\doibase
  10.1103/PhysRevD.94.034501} {\bibfield  {journal} {\bibinfo  {journal} {Phys.
  Rev. D}\ }\textbf {\bibinfo {volume} {94}},\ \bibinfo {pages} {034501}
  (\bibinfo {year} {2016})}\BibitemShut {NoStop}%
\bibitem [{\citenamefont {Guo}\ \emph {et~al.}(2018{\natexlab{a}})\citenamefont
  {Guo}, \citenamefont {Alexandru}, \citenamefont {Molina}, \citenamefont
  {Mai},\ and\ \citenamefont {D{\"o}ring}}]{Guo:2018zss}%
  \BibitemOpen
  \bibfield  {author} {\bibinfo {author} {\bibfnamefont {D.}~\bibnamefont
  {Guo}}, \bibinfo {author} {\bibfnamefont {A.}~\bibnamefont {Alexandru}},
  \bibinfo {author} {\bibfnamefont {R.}~\bibnamefont {Molina}}, \bibinfo
  {author} {\bibfnamefont {M.}~\bibnamefont {Mai}}, \ and\ \bibinfo {author}
  {\bibfnamefont {M.}~\bibnamefont {D{\"o}ring}},\ }\href {\doibase
  10.1103/PhysRevD.98.014507} {\bibfield  {journal} {\bibinfo  {journal} {Phys.
  Rev. D}\ }\textbf {\bibinfo {volume} {98}},\ \bibinfo {pages} {014507}
  (\bibinfo {year} {2018}{\natexlab{a}})},\ \Eprint
  {http://arxiv.org/abs/1803.02897} {arXiv:1803.02897} \BibitemShut {NoStop}%
\bibitem [{\citenamefont {Culver}\ \emph {et~al.}(2019)\citenamefont {Culver},
  \citenamefont {Mai}, \citenamefont {Alexandru}, \citenamefont {D{\"o}ring},\
  and\ \citenamefont {Lee}}]{Culver:2019qtx}%
  \BibitemOpen
  \bibfield  {author} {\bibinfo {author} {\bibfnamefont {C.}~\bibnamefont
  {Culver}}, \bibinfo {author} {\bibfnamefont {M.}~\bibnamefont {Mai}},
  \bibinfo {author} {\bibfnamefont {A.}~\bibnamefont {Alexandru}}, \bibinfo
  {author} {\bibfnamefont {M.}~\bibnamefont {D{\"o}ring}}, \ and\ \bibinfo
  {author} {\bibfnamefont {F.~X.}\ \bibnamefont {Lee}},\ }\href {\doibase
  10.1103/PhysRevD.100.034509} {\bibfield  {journal} {\bibinfo  {journal}
  {Phys. Rev. D}\ }\textbf {\bibinfo {volume} {100}},\ \bibinfo {pages}
  {034509} (\bibinfo {year} {2019})},\ \Eprint
  {http://arxiv.org/abs/1905.10202} {arXiv:1905.10202 [hep-lat]} \BibitemShut
  {NoStop}%
\bibitem [{\citenamefont {Culver}\ \emph {et~al.}(2020)\citenamefont {Culver},
  \citenamefont {Mai}, \citenamefont {Brett}, \citenamefont {Alexandru},\ and\
  \citenamefont {D{\"o}ring}}]{Culver:2019vvu}%
  \BibitemOpen
  \bibfield  {author} {\bibinfo {author} {\bibfnamefont {C.}~\bibnamefont
  {Culver}}, \bibinfo {author} {\bibfnamefont {M.}~\bibnamefont {Mai}},
  \bibinfo {author} {\bibfnamefont {R.}~\bibnamefont {Brett}}, \bibinfo
  {author} {\bibfnamefont {A.}~\bibnamefont {Alexandru}}, \ and\ \bibinfo
  {author} {\bibfnamefont {M.}~\bibnamefont {D{\"o}ring}},\ }\href {\doibase
  10.1103/PhysRevD.101.114507} {\bibfield  {journal} {\bibinfo  {journal}
  {Phys. Rev. D}\ }\textbf {\bibinfo {volume} {101}},\ \bibinfo {pages}
  {114507} (\bibinfo {year} {2020})}\BibitemShut {NoStop}%
\bibitem [{\citenamefont {Rummukainen}\ and\ \citenamefont
  {Gottlieb}(1995)}]{Rummukainen:1995vs}%
  \BibitemOpen
  \bibfield  {author} {\bibinfo {author} {\bibfnamefont {K.}~\bibnamefont
  {Rummukainen}}\ and\ \bibinfo {author} {\bibfnamefont {S.}~\bibnamefont
  {Gottlieb}},\ }\href {\doibase 10.1016/0550-3213(95)00313-H} {\bibfield
  {journal} {\bibinfo  {journal} {Nuclear Physics B}\ }\textbf {\bibinfo
  {volume} {450}},\ \bibinfo {pages} {397} (\bibinfo {year} {1995})},\ \Eprint
  {http://arxiv.org/abs/hep-lat/9503028} {arXiv:hep-lat/9503028} \BibitemShut
  {NoStop}%
\bibitem [{\citenamefont {Kim}\ \emph {et~al.}(2005)\citenamefont {Kim},
  \citenamefont {Sachrajda},\ and\ \citenamefont {Sharpe}}]{Kim:2005gf}%
  \BibitemOpen
  \bibfield  {author} {\bibinfo {author} {\bibfnamefont {C.}~\bibnamefont
  {Kim}}, \bibinfo {author} {\bibfnamefont {C.}~\bibnamefont {Sachrajda}}, \
  and\ \bibinfo {author} {\bibfnamefont {S.~R.}\ \bibnamefont {Sharpe}},\
  }\href {\doibase 10.1016/j.nuclphysb.2005.08.029} {\bibfield  {journal}
  {\bibinfo  {journal} {Nuclear Physics B}\ }\textbf {\bibinfo {volume}
  {727}},\ \bibinfo {pages} {218} (\bibinfo {year} {2005})},\ \Eprint
  {http://arxiv.org/abs/hep-lat/0507006} {arXiv:hep-lat/0507006} \BibitemShut
  {NoStop}%
\bibitem [{\citenamefont {G{\"o}ckeler}\ \emph {et~al.}(2012)\citenamefont
  {G{\"o}ckeler}, \citenamefont {Horsley}, \citenamefont {Lage}, \citenamefont
  {Mei{\ss}ner}, \citenamefont {Rakow}, \citenamefont {Rusetsky}, \citenamefont
  {Schierholz},\ and\ \citenamefont {Zanotti}}]{Gockeler:2012yj}%
  \BibitemOpen
  \bibfield  {author} {\bibinfo {author} {\bibfnamefont {M.}~\bibnamefont
  {G{\"o}ckeler}}, \bibinfo {author} {\bibfnamefont {R.}~\bibnamefont
  {Horsley}}, \bibinfo {author} {\bibfnamefont {M.}~\bibnamefont {Lage}},
  \bibinfo {author} {\bibfnamefont {U.-G.}\ \bibnamefont {Mei{\ss}ner}},
  \bibinfo {author} {\bibfnamefont {P.~E.~L.}\ \bibnamefont {Rakow}}, \bibinfo
  {author} {\bibfnamefont {A.}~\bibnamefont {Rusetsky}}, \bibinfo {author}
  {\bibfnamefont {G.}~\bibnamefont {Schierholz}}, \ and\ \bibinfo {author}
  {\bibfnamefont {J.~M.}\ \bibnamefont {Zanotti}},\ }\href {\doibase
  10.1103/PhysRevD.86.094513} {\bibfield  {journal} {\bibinfo  {journal} {Phys.
  Rev. D}\ }\textbf {\bibinfo {volume} {86}},\ \bibinfo {pages} {094513}
  (\bibinfo {year} {2012})},\ \Eprint {http://arxiv.org/abs/1206.4141}
  {arXiv:1206.4141 [hep-lat]} \BibitemShut {NoStop}%
\bibitem [{\citenamefont {Davoudi}\ and\ \citenamefont
  {Savage}(2011)}]{Davoudi:2011md}%
  \BibitemOpen
  \bibfield  {author} {\bibinfo {author} {\bibfnamefont {Z.}~\bibnamefont
  {Davoudi}}\ and\ \bibinfo {author} {\bibfnamefont {M.~J.}\ \bibnamefont
  {Savage}},\ }\href {\doibase 10.1103/PhysRevD.84.114502} {\bibfield
  {journal} {\bibinfo  {journal} {Phys. Rev.}\ }\textbf {\bibinfo {volume}
  {D84}},\ \bibinfo {pages} {114502} (\bibinfo {year} {2011})},\ \Eprint
  {http://arxiv.org/abs/1108.5371} {arXiv:1108.5371 [hep-lat]} \BibitemShut
  {NoStop}%
\bibitem [{\citenamefont {Fu}(2012)}]{Fu:2011xz}%
  \BibitemOpen
  \bibfield  {author} {\bibinfo {author} {\bibfnamefont {Z.}~\bibnamefont
  {Fu}},\ }\href {\doibase 10.1103/PhysRevD.85.014506} {\bibfield  {journal}
  {\bibinfo  {journal} {Phys. Rev.}\ }\textbf {\bibinfo {volume} {D85}},\
  \bibinfo {pages} {014506} (\bibinfo {year} {2012})},\ \Eprint
  {http://arxiv.org/abs/1110.0319} {arXiv:1110.0319 [hep-lat]} \BibitemShut
  {NoStop}%
\bibitem [{\citenamefont {Leskovec}\ and\ \citenamefont
  {Prelovsek}(2012)}]{Leskovec:2012gb}%
  \BibitemOpen
  \bibfield  {author} {\bibinfo {author} {\bibfnamefont {L.}~\bibnamefont
  {Leskovec}}\ and\ \bibinfo {author} {\bibfnamefont {S.}~\bibnamefont
  {Prelovsek}},\ }\href {\doibase 10.1103/PhysRevD.85.114507} {\bibfield
  {journal} {\bibinfo  {journal} {Phys. Rev. D}\ }\textbf {\bibinfo {volume}
  {85}},\ \bibinfo {pages} {114507} (\bibinfo {year} {2012})}\BibitemShut
  {NoStop}%
\bibitem [{\citenamefont {Li}\ \emph {et~al.}(2021)\citenamefont {Li},
  \citenamefont {Wu}, \citenamefont {Young},\ and\ \citenamefont
  {Lee}}]{Li:2021General}%
  \BibitemOpen
  \bibfield  {author} {\bibinfo {author} {\bibfnamefont {Y.}~\bibnamefont
  {Li}}, \bibinfo {author} {\bibfnamefont {J.-J.}\ \bibnamefont {Wu}}, \bibinfo
  {author} {\bibfnamefont {R.~D.}\ \bibnamefont {Young}}, \ and\ \bibinfo
  {author} {\bibfnamefont {T.-S.~H.}\ \bibnamefont {Lee}},\ }\href@noop {}
  {\bibfield  {journal} {\bibinfo  {journal} {In preparation}\ } (\bibinfo
  {year} {2021})}\BibitemShut {NoStop}%
\bibitem [{\citenamefont {Wu}\ \emph {et~al.}(2016)\citenamefont {Wu},
  \citenamefont {Lee}, \citenamefont {Leinweber}, \citenamefont {Thomas},\ and\
  \citenamefont {Young}}]{Wu:2015evh}%
  \BibitemOpen
  \bibfield  {author} {\bibinfo {author} {\bibfnamefont {J.-J.}\ \bibnamefont
  {Wu}}, \bibinfo {author} {\bibfnamefont {T.~S.~H.}\ \bibnamefont {Lee}},
  \bibinfo {author} {\bibfnamefont {D.~B.}\ \bibnamefont {Leinweber}}, \bibinfo
  {author} {\bibfnamefont {A.~W.}\ \bibnamefont {Thomas}}, \ and\ \bibinfo
  {author} {\bibfnamefont {R.~D.}\ \bibnamefont {Young}},\ }\href {\doibase
  10.7566/JPSCP.10.062002} {\bibfield  {journal} {\bibinfo  {journal} {JPS
  Conf. Proc.}\ }\textbf {\bibinfo {volume} {10}},\ \bibinfo {pages} {062002}
  (\bibinfo {year} {2016})},\ \Eprint {http://arxiv.org/abs/1512.02771}
  {arXiv:1512.02771 [hep-lat]} \BibitemShut {NoStop}%
\bibitem [{\citenamefont {Blanton}\ and\ \citenamefont
  {Sharpe}(2020)}]{Blanton:2020gha}%
  \BibitemOpen
  \bibfield  {author} {\bibinfo {author} {\bibfnamefont {T.~D.}\ \bibnamefont
  {Blanton}}\ and\ \bibinfo {author} {\bibfnamefont {S.~R.}\ \bibnamefont
  {Sharpe}},\ }\href {\doibase 10.1103/PhysRevD.102.054520} {\bibfield
  {journal} {\bibinfo  {journal} {Phys. Rev. D}\ }\textbf {\bibinfo {volume}
  {102}},\ \bibinfo {pages} {054520} (\bibinfo {year} {2020})},\ \Eprint
  {http://arxiv.org/abs/2007.16188} {arXiv:2007.16188} \BibitemShut {NoStop}%
\bibitem [{\citenamefont {Dudek}\ \emph {et~al.}(2012)\citenamefont {Dudek},
  \citenamefont {Edwards},\ and\ \citenamefont {Thomas}}]{Dudek:2012gj}%
  \BibitemOpen
  \bibfield  {author} {\bibinfo {author} {\bibfnamefont {J.~J.}\ \bibnamefont
  {Dudek}}, \bibinfo {author} {\bibfnamefont {R.~G.}\ \bibnamefont {Edwards}},
  \ and\ \bibinfo {author} {\bibfnamefont {C.~E.}\ \bibnamefont {Thomas}},\
  }\href {\doibase 10.1103/PhysRevD.86.034031} {\bibfield  {journal} {\bibinfo
  {journal} {Phys. Rev. D}\ }\textbf {\bibinfo {volume} {86}},\ \bibinfo
  {pages} {034031} (\bibinfo {year} {2012})},\ \Eprint
  {http://arxiv.org/abs/1203.6041} {arXiv:1203.6041 [hep-ph]} \BibitemShut
  {NoStop}%
\bibitem [{\citenamefont {McElvain}(2017)}]{McElvain:2017Harmonic}%
  \BibitemOpen
  \bibfield  {author} {\bibinfo {author} {\bibfnamefont {K.~S.}\ \bibnamefont
  {McElvain}},\ }\emph {\bibinfo {title} {Harmonic {{Oscillator Based Effective
  Theory}}, {{Connecting LQCD}} to {{Nuclear Structure}}}},\ \href@noop {}
  {Ph.D. thesis},\ \bibinfo  {school} {UC Berkeley} (\bibinfo {year}
  {2017})\BibitemShut {NoStop}%
\bibitem [{\citenamefont {McElvain}\ and\ \citenamefont
  {Haxton}(2019)}]{McElvain:2019ltw}%
  \BibitemOpen
  \bibfield  {author} {\bibinfo {author} {\bibfnamefont {K.}~\bibnamefont
  {McElvain}}\ and\ \bibinfo {author} {\bibfnamefont {W.}~\bibnamefont
  {Haxton}},\ }\href {\doibase 10.1016/j.physletb.2019.134880} {\bibfield
  {journal} {\bibinfo  {journal} {Physics Letters B}\ }\textbf {\bibinfo
  {volume} {797}},\ \bibinfo {pages} {134880} (\bibinfo {year}
  {2019})}\BibitemShut {NoStop}%
\bibitem [{\citenamefont {Drischler}\ \emph {et~al.}(2019)\citenamefont
  {Drischler}, \citenamefont {Haxton}, \citenamefont {McElvain}, \citenamefont
  {Mereghetti}, \citenamefont {Nicholson}, \citenamefont {Vranas},\ and\
  \citenamefont {{Walker-Loud}}}]{Drischler:2019xuo}%
  \BibitemOpen
  \bibfield  {author} {\bibinfo {author} {\bibfnamefont {C.}~\bibnamefont
  {Drischler}}, \bibinfo {author} {\bibfnamefont {W.}~\bibnamefont {Haxton}},
  \bibinfo {author} {\bibfnamefont {K.}~\bibnamefont {McElvain}}, \bibinfo
  {author} {\bibfnamefont {E.}~\bibnamefont {Mereghetti}}, \bibinfo {author}
  {\bibfnamefont {A.}~\bibnamefont {Nicholson}}, \bibinfo {author}
  {\bibfnamefont {P.}~\bibnamefont {Vranas}}, \ and\ \bibinfo {author}
  {\bibfnamefont {A.}~\bibnamefont {{Walker-Loud}}},\ }\href@noop {} {\bibfield
   {journal} {\bibinfo  {journal} {arXiv:1910.07961 [hep-ex, physics:hep-lat,
  physics:hep-ph, physics:nucl-ex, physics:nucl-th]}\ } (\bibinfo {year}
  {2019})},\ \Eprint {http://arxiv.org/abs/1910.07961} {arXiv:1910.07961
  [hep-ex, physics:hep-lat, physics:hep-ph, physics:nucl-ex, physics:nucl-th]}
  \BibitemShut {NoStop}%
\bibitem [{\citenamefont {Bernard}\ \emph {et~al.}(2011)\citenamefont
  {Bernard}, \citenamefont {Lage}, \citenamefont {Mei{\ss}ner},\ and\
  \citenamefont {Rusetsky}}]{Bernard:2010fp}%
  \BibitemOpen
  \bibfield  {author} {\bibinfo {author} {\bibfnamefont {V.}~\bibnamefont
  {Bernard}}, \bibinfo {author} {\bibfnamefont {M.}~\bibnamefont {Lage}},
  \bibinfo {author} {\bibfnamefont {U.-G.}\ \bibnamefont {Mei{\ss}ner}}, \ and\
  \bibinfo {author} {\bibfnamefont {A.}~\bibnamefont {Rusetsky}},\ }\href
  {\doibase 10.1007/JHEP01(2011)019} {\bibfield  {journal} {\bibinfo  {journal}
  {J. High Energ. Phys.}\ }\textbf {\bibinfo {volume} {2011}},\ \bibinfo
  {pages} {19} (\bibinfo {year} {2011})},\ \Eprint
  {http://arxiv.org/abs/1010.6018} {arXiv:1010.6018 [hep-lat]} \BibitemShut
  {NoStop}%
\bibitem [{\citenamefont {Guo}\ \emph {et~al.}(2013)\citenamefont {Guo},
  \citenamefont {Dudek}, \citenamefont {Edwards},\ and\ \citenamefont
  {Szczepaniak}}]{Guo:2012hv}%
  \BibitemOpen
  \bibfield  {author} {\bibinfo {author} {\bibfnamefont {P.}~\bibnamefont
  {Guo}}, \bibinfo {author} {\bibfnamefont {J.~J.}\ \bibnamefont {Dudek}},
  \bibinfo {author} {\bibfnamefont {R.~G.}\ \bibnamefont {Edwards}}, \ and\
  \bibinfo {author} {\bibfnamefont {A.~P.}\ \bibnamefont {Szczepaniak}},\
  }\href {\doibase 10.1103/PhysRevD.88.014501} {\bibfield  {journal} {\bibinfo
  {journal} {Phys. Rev. D}\ }\textbf {\bibinfo {volume} {88}},\ \bibinfo
  {pages} {014501} (\bibinfo {year} {2013})},\ \Eprint
  {http://arxiv.org/abs/1211.0929} {arXiv:1211.0929 [hep-lat]} \BibitemShut
  {NoStop}%
\bibitem [{\citenamefont {He}\ \emph {et~al.}(2005)\citenamefont {He},
  \citenamefont {Feng},\ and\ \citenamefont {Liu}}]{He:2005ey}%
  \BibitemOpen
  \bibfield  {author} {\bibinfo {author} {\bibfnamefont {S.}~\bibnamefont
  {He}}, \bibinfo {author} {\bibfnamefont {X.}~\bibnamefont {Feng}}, \ and\
  \bibinfo {author} {\bibfnamefont {C.}~\bibnamefont {Liu}},\ }\href {\doibase
  10.1088/1126-6708/2005/07/011} {\bibfield  {journal} {\bibinfo  {journal}
  {JHEP}\ }\textbf {\bibinfo {volume} {07}},\ \bibinfo {pages} {011} (\bibinfo
  {year} {2005})},\ \Eprint {http://arxiv.org/abs/hep-lat/0504019}
  {arXiv:hep-lat/0504019} \BibitemShut {NoStop}%
\bibitem [{\citenamefont {Hu}\ \emph {et~al.}(2016)\citenamefont {Hu},
  \citenamefont {Molina}, \citenamefont {D{\"o}ring},\ and\ \citenamefont
  {Alexandru}}]{Hu:2016shf}%
  \BibitemOpen
  \bibfield  {author} {\bibinfo {author} {\bibfnamefont {B.}~\bibnamefont
  {Hu}}, \bibinfo {author} {\bibfnamefont {R.}~\bibnamefont {Molina}}, \bibinfo
  {author} {\bibfnamefont {M.}~\bibnamefont {D{\"o}ring}}, \ and\ \bibinfo
  {author} {\bibfnamefont {A.}~\bibnamefont {Alexandru}},\ }\href {\doibase
  10.1103/PhysRevLett.117.122001} {\bibfield  {journal} {\bibinfo  {journal}
  {Phys. Rev. Lett.}\ }\textbf {\bibinfo {volume} {117}},\ \bibinfo {pages}
  {122001} (\bibinfo {year} {2016})},\ \Eprint
  {http://arxiv.org/abs/1605.04823} {arXiv:1605.04823} \BibitemShut {NoStop}%
\bibitem [{\citenamefont {Lage}\ \emph {et~al.}(2009)\citenamefont {Lage},
  \citenamefont {Meissner},\ and\ \citenamefont {Rusetsky}}]{Lage:2009zv}%
  \BibitemOpen
  \bibfield  {author} {\bibinfo {author} {\bibfnamefont {M.}~\bibnamefont
  {Lage}}, \bibinfo {author} {\bibfnamefont {U.-G.}\ \bibnamefont {Meissner}},
  \ and\ \bibinfo {author} {\bibfnamefont {A.}~\bibnamefont {Rusetsky}},\
  }\href {\doibase 10.1016/j.physletb.2009.10.055} {\bibfield  {journal}
  {\bibinfo  {journal} {Phys. Lett.}\ }\textbf {\bibinfo {volume} {B681}},\
  \bibinfo {pages} {439} (\bibinfo {year} {2009})},\ \Eprint
  {http://arxiv.org/abs/0905.0069} {arXiv:0905.0069 [hep-lat]} \BibitemShut
  {NoStop}%
\bibitem [{\citenamefont {Li}\ and\ \citenamefont {Liu}(2013)}]{Li:2012bi}%
  \BibitemOpen
  \bibfield  {author} {\bibinfo {author} {\bibfnamefont {N.}~\bibnamefont
  {Li}}\ and\ \bibinfo {author} {\bibfnamefont {C.}~\bibnamefont {Liu}},\
  }\href {\doibase 10.1103/PhysRevD.87.014502} {\bibfield  {journal} {\bibinfo
  {journal} {Phys. Rev. D}\ }\textbf {\bibinfo {volume} {87}},\ \bibinfo
  {pages} {014502} (\bibinfo {year} {2013})},\ \Eprint
  {http://arxiv.org/abs/1209.2201} {arXiv:1209.2201 [hep-lat]} \BibitemShut
  {NoStop}%
\bibitem [{\citenamefont {Beane}\ \emph {et~al.}(2004)\citenamefont {Beane},
  \citenamefont {Bedaque}, \citenamefont {Parreno},\ and\ \citenamefont
  {Savage}}]{Beane:2003da}%
  \BibitemOpen
  \bibfield  {author} {\bibinfo {author} {\bibfnamefont {S.~R.}\ \bibnamefont
  {Beane}}, \bibinfo {author} {\bibfnamefont {P.~F.}\ \bibnamefont {Bedaque}},
  \bibinfo {author} {\bibfnamefont {A.}~\bibnamefont {Parreno}}, \ and\
  \bibinfo {author} {\bibfnamefont {M.~J.}\ \bibnamefont {Savage}},\ }\href
  {\doibase 10.1016/j.physletb.2004.02.007} {\bibfield  {journal} {\bibinfo
  {journal} {Phys. Lett.}\ }\textbf {\bibinfo {volume} {B585}},\ \bibinfo
  {pages} {106} (\bibinfo {year} {2004})},\ \Eprint
  {http://arxiv.org/abs/hep-lat/0312004} {arXiv:hep-lat/0312004} \BibitemShut
  {NoStop}%
\bibitem [{\citenamefont {Beane}\ \emph {et~al.}(2005)\citenamefont {Beane},
  \citenamefont {Bedaque}, \citenamefont {Parreno},\ and\ \citenamefont
  {Savage}}]{Beane:2003yx}%
  \BibitemOpen
  \bibfield  {author} {\bibinfo {author} {\bibfnamefont {S.~R.}\ \bibnamefont
  {Beane}}, \bibinfo {author} {\bibfnamefont {P.~F.}\ \bibnamefont {Bedaque}},
  \bibinfo {author} {\bibfnamefont {A.}~\bibnamefont {Parreno}}, \ and\
  \bibinfo {author} {\bibfnamefont {M.~J.}\ \bibnamefont {Savage}},\ }\href
  {\doibase 10.1016/j.nuclphysa.2004.09.081} {\bibfield  {journal} {\bibinfo
  {journal} {Nucl. Phys.}\ }\textbf {\bibinfo {volume} {A747}},\ \bibinfo
  {pages} {55} (\bibinfo {year} {2005})},\ \Eprint
  {http://arxiv.org/abs/nucl-th/0311027} {arXiv:nucl-th/0311027} \BibitemShut
  {NoStop}%
\bibitem [{\citenamefont {Brice{\~n}o}\ \emph {et~al.}(2013)\citenamefont
  {Brice{\~n}o}, \citenamefont {Davoudi}, \citenamefont {Luu},\ and\
  \citenamefont {Savage}}]{Briceno:2013bda}%
  \BibitemOpen
  \bibfield  {author} {\bibinfo {author} {\bibfnamefont {R.~A.}\ \bibnamefont
  {Brice{\~n}o}}, \bibinfo {author} {\bibfnamefont {Z.}~\bibnamefont
  {Davoudi}}, \bibinfo {author} {\bibfnamefont {T.~C.}\ \bibnamefont {Luu}}, \
  and\ \bibinfo {author} {\bibfnamefont {M.~J.}\ \bibnamefont {Savage}},\
  }\href {\doibase 10.1103/PhysRevD.88.114507} {\bibfield  {journal} {\bibinfo
  {journal} {Phys. Rev. D}\ }\textbf {\bibinfo {volume} {88}},\ \bibinfo
  {pages} {114507} (\bibinfo {year} {2013})},\ \Eprint
  {http://arxiv.org/abs/1309.3556} {arXiv:1309.3556 [hep-lat]} \BibitemShut
  {NoStop}%
\bibitem [{\citenamefont {Meng}\ \emph {et~al.}(2004)\citenamefont {Meng},
  \citenamefont {Miao}, \citenamefont {Du},\ and\ \citenamefont
  {Liu}}]{Meng:2003gm}%
  \BibitemOpen
  \bibfield  {author} {\bibinfo {author} {\bibfnamefont {G.}~\bibnamefont
  {Meng}}, \bibinfo {author} {\bibfnamefont {C.}~\bibnamefont {Miao}}, \bibinfo
  {author} {\bibfnamefont {X.}~\bibnamefont {Du}}, \ and\ \bibinfo {author}
  {\bibfnamefont {C.}~\bibnamefont {Liu}},\ }\href {\doibase
  10.1142/S0217751X04019627} {\bibfield  {journal} {\bibinfo  {journal} {Int.
  J. Mod. Phys. A}\ }\textbf {\bibinfo {volume} {19}},\ \bibinfo {pages} {4401}
  (\bibinfo {year} {2004})},\ \Eprint {http://arxiv.org/abs/hep-lat/0309048}
  {arXiv:hep-lat/0309048} \BibitemShut {NoStop}%
\bibitem [{\citenamefont {Bedaque}\ and\ \citenamefont
  {Chen}(2005)}]{Bedaque:2004ax}%
  \BibitemOpen
  \bibfield  {author} {\bibinfo {author} {\bibfnamefont {P.~F.}\ \bibnamefont
  {Bedaque}}\ and\ \bibinfo {author} {\bibfnamefont {J.-W.}\ \bibnamefont
  {Chen}},\ }\href {\doibase 10.1016/j.physletb.2005.04.045} {\bibfield
  {journal} {\bibinfo  {journal} {Physics Letters B}\ }\textbf {\bibinfo
  {volume} {616}},\ \bibinfo {pages} {208} (\bibinfo {year} {2005})},\ \Eprint
  {http://arxiv.org/abs/hep-lat/0412023} {arXiv:hep-lat/0412023} \BibitemShut
  {NoStop}%
\bibitem [{\citenamefont {Bedaque}(2004)}]{Bedaque:2004kc}%
  \BibitemOpen
  \bibfield  {author} {\bibinfo {author} {\bibfnamefont {P.~F.}\ \bibnamefont
  {Bedaque}},\ }\href {\doibase 10.1016/j.physletb.2004.04.045} {\bibfield
  {journal} {\bibinfo  {journal} {Physics Letters B}\ }\textbf {\bibinfo
  {volume} {593}},\ \bibinfo {pages} {82} (\bibinfo {year} {2004})},\ \Eprint
  {http://arxiv.org/abs/nucl-th/0402051} {arXiv:nucl-th/0402051} \BibitemShut
  {NoStop}%
\bibitem [{\citenamefont {{de Divitiis}}\ \emph {et~al.}(2004)\citenamefont
  {{de Divitiis}}, \citenamefont {Petronzio},\ and\ \citenamefont
  {Tantalo}}]{deDivitiis:2004kq}%
  \BibitemOpen
  \bibfield  {author} {\bibinfo {author} {\bibfnamefont {G.~M.}\ \bibnamefont
  {{de Divitiis}}}, \bibinfo {author} {\bibfnamefont {R.}~\bibnamefont
  {Petronzio}}, \ and\ \bibinfo {author} {\bibfnamefont {N.}~\bibnamefont
  {Tantalo}},\ }\href {\doibase 10.1016/j.physletb.2004.06.035} {\bibfield
  {journal} {\bibinfo  {journal} {Physics Letters B}\ }\textbf {\bibinfo
  {volume} {595}},\ \bibinfo {pages} {408} (\bibinfo {year} {2004})},\ \Eprint
  {http://arxiv.org/abs/hep-lat/0405002} {arXiv:hep-lat/0405002} \BibitemShut
  {NoStop}%
\bibitem [{\citenamefont {Sachrajda}\ and\ \citenamefont
  {Villadoro}(2005)}]{Sachrajda:2004mi}%
  \BibitemOpen
  \bibfield  {author} {\bibinfo {author} {\bibfnamefont {C.~T.}\ \bibnamefont
  {Sachrajda}}\ and\ \bibinfo {author} {\bibfnamefont {G.}~\bibnamefont
  {Villadoro}},\ }\href {\doibase 10.1016/j.physletb.2005.01.033} {\bibfield
  {journal} {\bibinfo  {journal} {Physics Letters B}\ }\textbf {\bibinfo
  {volume} {609}},\ \bibinfo {pages} {73} (\bibinfo {year} {2005})},\ \Eprint
  {http://arxiv.org/abs/hep-lat/0411033} {arXiv:hep-lat/0411033} \BibitemShut
  {NoStop}%
\bibitem [{\citenamefont {Polejaeva}\ and\ \citenamefont
  {Rusetsky}(2012)}]{Polejaeva:2012ut}%
  \BibitemOpen
  \bibfield  {author} {\bibinfo {author} {\bibfnamefont {K.}~\bibnamefont
  {Polejaeva}}\ and\ \bibinfo {author} {\bibfnamefont {A.}~\bibnamefont
  {Rusetsky}},\ }\href {\doibase 10.1140/epja/i2012-12067-8} {\bibfield
  {journal} {\bibinfo  {journal} {Eur. Phys. J. A}\ }\textbf {\bibinfo {volume}
  {48}},\ \bibinfo {pages} {67} (\bibinfo {year} {2012})},\ \Eprint
  {http://arxiv.org/abs/1203.1241} {arXiv:1203.1241 [hep-lat]} \BibitemShut
  {NoStop}%
\bibitem [{\citenamefont {Brice{\~n}o}\ and\ \citenamefont
  {Davoudi}(2013)}]{Briceno:2012rv}%
  \BibitemOpen
  \bibfield  {author} {\bibinfo {author} {\bibfnamefont {R.~A.}\ \bibnamefont
  {Brice{\~n}o}}\ and\ \bibinfo {author} {\bibfnamefont {Z.}~\bibnamefont
  {Davoudi}},\ }\href {\doibase 10.1103/PhysRevD.87.094507} {\bibfield
  {journal} {\bibinfo  {journal} {Phys. Rev. D}\ }\textbf {\bibinfo {volume}
  {87}},\ \bibinfo {pages} {094507} (\bibinfo {year} {2013})},\ \Eprint
  {http://arxiv.org/abs/1212.3398} {arXiv:1212.3398 [hep-lat]} \BibitemShut
  {NoStop}%
\bibitem [{\citenamefont {Hansen}\ and\ \citenamefont
  {Sharpe}(2014)}]{Hansen:2014eka}%
  \BibitemOpen
  \bibfield  {author} {\bibinfo {author} {\bibfnamefont {M.~T.}\ \bibnamefont
  {Hansen}}\ and\ \bibinfo {author} {\bibfnamefont {S.~R.}\ \bibnamefont
  {Sharpe}},\ }\href {\doibase 10.1103/PhysRevD.90.116003} {\bibfield
  {journal} {\bibinfo  {journal} {Phys. Rev. D}\ }\textbf {\bibinfo {volume}
  {90}},\ \bibinfo {pages} {116003} (\bibinfo {year} {2014})},\ \Eprint
  {http://arxiv.org/abs/1408.5933} {arXiv:1408.5933 [hep-lat]} \BibitemShut
  {NoStop}%
\bibitem [{\citenamefont {Hansen}\ and\ \citenamefont
  {Sharpe}(2015)}]{Hansen:2015zga}%
  \BibitemOpen
  \bibfield  {author} {\bibinfo {author} {\bibfnamefont {M.~T.}\ \bibnamefont
  {Hansen}}\ and\ \bibinfo {author} {\bibfnamefont {S.~R.}\ \bibnamefont
  {Sharpe}},\ }\href {\doibase 10.1103/PhysRevD.92.114509} {\bibfield
  {journal} {\bibinfo  {journal} {Phys. Rev. D}\ }\textbf {\bibinfo {volume}
  {92}},\ \bibinfo {pages} {114509} (\bibinfo {year} {2015})},\ \Eprint
  {http://arxiv.org/abs/1504.04248} {arXiv:1504.04248 [hep-lat]} \BibitemShut
  {NoStop}%
\bibitem [{\citenamefont {Brice{\~n}o}\ \emph {et~al.}(2017)\citenamefont
  {Brice{\~n}o}, \citenamefont {Hansen},\ and\ \citenamefont
  {Sharpe}}]{Briceno:2017tce}%
  \BibitemOpen
  \bibfield  {author} {\bibinfo {author} {\bibfnamefont {R.~A.}\ \bibnamefont
  {Brice{\~n}o}}, \bibinfo {author} {\bibfnamefont {M.~T.}\ \bibnamefont
  {Hansen}}, \ and\ \bibinfo {author} {\bibfnamefont {S.~R.}\ \bibnamefont
  {Sharpe}},\ }\href {\doibase 10.1103/PhysRevD.95.074510} {\bibfield
  {journal} {\bibinfo  {journal} {Phys. Rev. D}\ }\textbf {\bibinfo {volume}
  {95}},\ \bibinfo {pages} {074510} (\bibinfo {year} {2017})},\ \Eprint
  {http://arxiv.org/abs/1701.07465} {arXiv:1701.07465 [hep-lat]} \BibitemShut
  {NoStop}%
\bibitem [{\citenamefont {Hammer}\ \emph
  {et~al.}(2017{\natexlab{a}})\citenamefont {Hammer}, \citenamefont {Pang},\
  and\ \citenamefont {Rusetsky}}]{Hammer:2017kms}%
  \BibitemOpen
  \bibfield  {author} {\bibinfo {author} {\bibfnamefont {H.~W.}\ \bibnamefont
  {Hammer}}, \bibinfo {author} {\bibfnamefont {J.~Y.}\ \bibnamefont {Pang}}, \
  and\ \bibinfo {author} {\bibfnamefont {A.}~\bibnamefont {Rusetsky}},\ }\href
  {\doibase 10.1007/JHEP10(2017)115} {\bibfield  {journal} {\bibinfo  {journal}
  {JHEP}\ }\textbf {\bibinfo {volume} {10}},\ \bibinfo {pages} {115} (\bibinfo
  {year} {2017}{\natexlab{a}})},\ \Eprint {http://arxiv.org/abs/1707.02176}
  {arXiv:1707.02176 [hep-lat]} \BibitemShut {NoStop}%
\bibitem [{\citenamefont {Hammer}\ \emph
  {et~al.}(2017{\natexlab{b}})\citenamefont {Hammer}, \citenamefont {Pang},\
  and\ \citenamefont {Rusetsky}}]{Hammer:2017uqm}%
  \BibitemOpen
  \bibfield  {author} {\bibinfo {author} {\bibfnamefont {H.-W.}\ \bibnamefont
  {Hammer}}, \bibinfo {author} {\bibfnamefont {J.-Y.}\ \bibnamefont {Pang}}, \
  and\ \bibinfo {author} {\bibfnamefont {A.}~\bibnamefont {Rusetsky}},\ }\href
  {\doibase 10.1007/JHEP09(2017)109} {\bibfield  {journal} {\bibinfo  {journal}
  {JHEP}\ }\textbf {\bibinfo {volume} {09}},\ \bibinfo {pages} {109} (\bibinfo
  {year} {2017}{\natexlab{b}})},\ \Eprint {http://arxiv.org/abs/1706.07700}
  {arXiv:1706.07700 [hep-lat]} \BibitemShut {NoStop}%
\bibitem [{\citenamefont {Hansen}\ \emph {et~al.}(2017)\citenamefont {Hansen},
  \citenamefont {Meyer},\ and\ \citenamefont {Robaina}}]{Hansen:2017mnd}%
  \BibitemOpen
  \bibfield  {author} {\bibinfo {author} {\bibfnamefont {M.~T.}\ \bibnamefont
  {Hansen}}, \bibinfo {author} {\bibfnamefont {H.~B.}\ \bibnamefont {Meyer}}, \
  and\ \bibinfo {author} {\bibfnamefont {D.}~\bibnamefont {Robaina}},\ }\href
  {\doibase 10.1103/PhysRevD.96.094513} {\bibfield  {journal} {\bibinfo
  {journal} {Phys. Rev. D}\ }\textbf {\bibinfo {volume} {96}},\ \bibinfo
  {pages} {094513} (\bibinfo {year} {2017})}\BibitemShut {NoStop}%
\bibitem [{\citenamefont {Mai}\ and\ \citenamefont
  {D{\"o}ring}(2017)}]{Mai:2017bge}%
  \BibitemOpen
  \bibfield  {author} {\bibinfo {author} {\bibfnamefont {M.}~\bibnamefont
  {Mai}}\ and\ \bibinfo {author} {\bibfnamefont {M.}~\bibnamefont
  {D{\"o}ring}},\ }\href {\doibase 10.1140/epja/i2017-12440-1} {\bibfield
  {journal} {\bibinfo  {journal} {Eur. Phys. J. A}\ }\textbf {\bibinfo {volume}
  {53}},\ \bibinfo {pages} {240} (\bibinfo {year} {2017})}\BibitemShut
  {NoStop}%
\bibitem [{\citenamefont {D{\"o}ring}\ \emph {et~al.}(2018)\citenamefont
  {D{\"o}ring}, \citenamefont {Hammer}, \citenamefont {Mai}, \citenamefont
  {Pang}, \citenamefont {Rusetsky},\ and\ \citenamefont {Wu}}]{Doring:2018xxx}%
  \BibitemOpen
  \bibfield  {author} {\bibinfo {author} {\bibfnamefont {M.}~\bibnamefont
  {D{\"o}ring}}, \bibinfo {author} {\bibfnamefont {H.-W.}\ \bibnamefont
  {Hammer}}, \bibinfo {author} {\bibfnamefont {M.}~\bibnamefont {Mai}},
  \bibinfo {author} {\bibfnamefont {J.-Y.}\ \bibnamefont {Pang}}, \bibinfo
  {author} {\bibfnamefont {A.}~\bibnamefont {Rusetsky}}, \ and\ \bibinfo
  {author} {\bibfnamefont {J.}~\bibnamefont {Wu}},\ }\href {\doibase
  10.1103/PhysRevD.97.114508} {\bibfield  {journal} {\bibinfo  {journal} {Phys.
  Rev. D}\ }\textbf {\bibinfo {volume} {97}},\ \bibinfo {pages} {114508}
  (\bibinfo {year} {2018})}\BibitemShut {NoStop}%
\bibitem [{\citenamefont {Guo}\ \emph {et~al.}(2018{\natexlab{b}})\citenamefont
  {Guo}, \citenamefont {D{\"o}ring},\ and\ \citenamefont
  {Szczepaniak}}]{Guo:2018ibd}%
  \BibitemOpen
  \bibfield  {author} {\bibinfo {author} {\bibfnamefont {P.}~\bibnamefont
  {Guo}}, \bibinfo {author} {\bibfnamefont {M.}~\bibnamefont {D{\"o}ring}}, \
  and\ \bibinfo {author} {\bibfnamefont {A.~P.}\ \bibnamefont {Szczepaniak}},\
  }\href {\doibase 10.1103/PhysRevD.98.094502} {\bibfield  {journal} {\bibinfo
  {journal} {Phys. Rev.}\ }\textbf {\bibinfo {volume} {D98}},\ \bibinfo {pages}
  {094502} (\bibinfo {year} {2018}{\natexlab{b}})},\ \Eprint
  {http://arxiv.org/abs/1810.01261} {arXiv:1810.01261 [hep-lat]} \BibitemShut
  {NoStop}%
\bibitem [{\citenamefont {Mai}\ and\ \citenamefont
  {D{\"o}ring}(2019)}]{Mai:2018djl}%
  \BibitemOpen
  \bibfield  {author} {\bibinfo {author} {\bibfnamefont {M.}~\bibnamefont
  {Mai}}\ and\ \bibinfo {author} {\bibfnamefont {M.}~\bibnamefont
  {D{\"o}ring}},\ }\href {\doibase 10.1103/PhysRevLett.122.062503} {\bibfield
  {journal} {\bibinfo  {journal} {Phys. Rev. Lett.}\ }\textbf {\bibinfo
  {volume} {122}},\ \bibinfo {pages} {062503} (\bibinfo {year}
  {2019})}\BibitemShut {NoStop}%
\bibitem [{\citenamefont {Meng}\ \emph {et~al.}(2018)\citenamefont {Meng},
  \citenamefont {Liu}, \citenamefont {Mei{\ss}ner},\ and\ \citenamefont
  {Rusetsky}}]{Meng:2017jgx}%
  \BibitemOpen
  \bibfield  {author} {\bibinfo {author} {\bibfnamefont {Y.}~\bibnamefont
  {Meng}}, \bibinfo {author} {\bibfnamefont {C.}~\bibnamefont {Liu}}, \bibinfo
  {author} {\bibfnamefont {U.-G.}\ \bibnamefont {Mei{\ss}ner}}, \ and\ \bibinfo
  {author} {\bibfnamefont {A.}~\bibnamefont {Rusetsky}},\ }\href {\doibase
  10.1103/PhysRevD.98.014508} {\bibfield  {journal} {\bibinfo  {journal} {Phys.
  Rev. D}\ }\textbf {\bibinfo {volume} {98}},\ \bibinfo {pages} {014508}
  (\bibinfo {year} {2018})},\ \Eprint {http://arxiv.org/abs/1712.08464}
  {arXiv:1712.08464 [hep-lat]} \BibitemShut {NoStop}%
\bibitem [{\citenamefont {Bulava}\ and\ \citenamefont
  {Hansen}(2019)}]{Bulava:2019kbi}%
  \BibitemOpen
  \bibfield  {author} {\bibinfo {author} {\bibfnamefont {J.}~\bibnamefont
  {Bulava}}\ and\ \bibinfo {author} {\bibfnamefont {M.~T.}\ \bibnamefont
  {Hansen}},\ }\href {\doibase 10.1103/PhysRevD.100.034521} {\bibfield
  {journal} {\bibinfo  {journal} {Phys. Rev. D}\ }\textbf {\bibinfo {volume}
  {100}},\ \bibinfo {pages} {034521} (\bibinfo {year} {2019})}\BibitemShut
  {NoStop}%
\bibitem [{\citenamefont {Hansen}\ and\ \citenamefont
  {Sharpe}(2019)}]{Hansen:2019nir}%
  \BibitemOpen
  \bibfield  {author} {\bibinfo {author} {\bibfnamefont {M.~T.}\ \bibnamefont
  {Hansen}}\ and\ \bibinfo {author} {\bibfnamefont {S.~R.}\ \bibnamefont
  {Sharpe}},\ }\href {\doibase 10.1146/annurev-nucl-101918-023723} {\bibfield
  {journal} {\bibinfo  {journal} {Annu. Rev. Nucl. Part. Sci.}\ }\textbf
  {\bibinfo {volume} {69}},\ \bibinfo {pages} {65} (\bibinfo {year}
  {2019})}\BibitemShut {NoStop}%
\bibitem [{\citenamefont {Jackura}\ \emph {et~al.}(2019)\citenamefont
  {Jackura}, \citenamefont {Dawid}, \citenamefont
  {{Fern{\'a}ndez-Ram{\'i}rez}}, \citenamefont {Mathieu}, \citenamefont
  {Mikhasenko}, \citenamefont {Pilloni}, \citenamefont {Sharpe},\ and\
  \citenamefont {Szczepaniak}}]{Jackura:2019bmu}%
  \BibitemOpen
  \bibfield  {author} {\bibinfo {author} {\bibfnamefont {A.~W.}\ \bibnamefont
  {Jackura}}, \bibinfo {author} {\bibfnamefont {S.~M.}\ \bibnamefont {Dawid}},
  \bibinfo {author} {\bibfnamefont {C.}~\bibnamefont
  {{Fern{\'a}ndez-Ram{\'i}rez}}}, \bibinfo {author} {\bibfnamefont
  {V.}~\bibnamefont {Mathieu}}, \bibinfo {author} {\bibfnamefont
  {M.}~\bibnamefont {Mikhasenko}}, \bibinfo {author} {\bibfnamefont
  {A.}~\bibnamefont {Pilloni}}, \bibinfo {author} {\bibfnamefont {S.~R.}\
  \bibnamefont {Sharpe}}, \ and\ \bibinfo {author} {\bibfnamefont {A.~P.}\
  \bibnamefont {Szczepaniak}},\ }\href {\doibase 10.1103/PhysRevD.100.034508}
  {\bibfield  {journal} {\bibinfo  {journal} {Phys. Rev. D}\ }\textbf {\bibinfo
  {volume} {100}},\ \bibinfo {pages} {034508} (\bibinfo {year}
  {2019})}\BibitemShut {NoStop}%
\bibitem [{\citenamefont {Pang}\ \emph {et~al.}(2019)\citenamefont {Pang},
  \citenamefont {Wu}, \citenamefont {Hammer}, \citenamefont {Mei{\ss}ner},\
  and\ \citenamefont {Rusetsky}}]{Pang:2019dfe}%
  \BibitemOpen
  \bibfield  {author} {\bibinfo {author} {\bibfnamefont {J.-Y.}\ \bibnamefont
  {Pang}}, \bibinfo {author} {\bibfnamefont {J.-J.}\ \bibnamefont {Wu}},
  \bibinfo {author} {\bibfnamefont {H.-W.}\ \bibnamefont {Hammer}}, \bibinfo
  {author} {\bibfnamefont {U.-G.}\ \bibnamefont {Mei{\ss}ner}}, \ and\ \bibinfo
  {author} {\bibfnamefont {A.}~\bibnamefont {Rusetsky}},\ }\href {\doibase
  10.1103/PhysRevD.99.074513} {\bibfield  {journal} {\bibinfo  {journal} {Phys.
  Rev. D}\ }\textbf {\bibinfo {volume} {99}},\ \bibinfo {pages} {074513}
  (\bibinfo {year} {2019})},\ \Eprint {http://arxiv.org/abs/1902.01111}
  {arXiv:1902.01111 [hep-lat]} \BibitemShut {NoStop}%
\bibitem [{\citenamefont {{Romero-L{\'o}pez}}\ \emph
  {et~al.}(2019)\citenamefont {{Romero-L{\'o}pez}}, \citenamefont {Sharpe},
  \citenamefont {Blanton}, \citenamefont {Brice{\~n}o},\ and\ \citenamefont
  {Hansen}}]{Romero-Lopez:2019qrt}%
  \BibitemOpen
  \bibfield  {author} {\bibinfo {author} {\bibfnamefont {F.}~\bibnamefont
  {{Romero-L{\'o}pez}}}, \bibinfo {author} {\bibfnamefont {S.~R.}\ \bibnamefont
  {Sharpe}}, \bibinfo {author} {\bibfnamefont {T.~D.}\ \bibnamefont {Blanton}},
  \bibinfo {author} {\bibfnamefont {R.~A.}\ \bibnamefont {Brice{\~n}o}}, \ and\
  \bibinfo {author} {\bibfnamefont {M.~T.}\ \bibnamefont {Hansen}},\ }\href
  {\doibase 10.1007/JHEP10(2019)007} {\bibfield  {journal} {\bibinfo  {journal}
  {JHEP}\ }\textbf {\bibinfo {volume} {10}},\ \bibinfo {pages} {007} (\bibinfo
  {year} {2019})},\ \Eprint {http://arxiv.org/abs/1908.02411} {arXiv:1908.02411
  [hep-lat]} \BibitemShut {NoStop}%
\bibitem [{\citenamefont {Blanton}\ \emph {et~al.}(2020)\citenamefont
  {Blanton}, \citenamefont {{Romero-L{\'o}pez}},\ and\ \citenamefont
  {Sharpe}}]{Blanton:2019vdk}%
  \BibitemOpen
  \bibfield  {author} {\bibinfo {author} {\bibfnamefont {T.~D.}\ \bibnamefont
  {Blanton}}, \bibinfo {author} {\bibfnamefont {F.}~\bibnamefont
  {{Romero-L{\'o}pez}}}, \ and\ \bibinfo {author} {\bibfnamefont {S.~R.}\
  \bibnamefont {Sharpe}},\ }\href {\doibase 10.1103/PhysRevLett.124.032001}
  {\bibfield  {journal} {\bibinfo  {journal} {Phys. Rev. Lett.}\ }\textbf
  {\bibinfo {volume} {124}},\ \bibinfo {pages} {032001} (\bibinfo {year}
  {2020})},\ \Eprint {http://arxiv.org/abs/1909.02973} {arXiv:1909.02973}
  \BibitemShut {NoStop}%
\bibitem [{\citenamefont {{for the Hadron Spectrum Collaboration}}\ \emph
  {et~al.}(2021)\citenamefont {{for the Hadron Spectrum Collaboration}},
  \citenamefont {Hansen}, \citenamefont {Brice{\~n}o}, \citenamefont {Edwards},
  \citenamefont {Thomas},\ and\ \citenamefont {Wilson}}]{Hansen:2020otl}%
  \BibitemOpen
  \bibfield  {author} {\bibinfo {author} {\bibnamefont {{for the Hadron
  Spectrum Collaboration}}}, \bibinfo {author} {\bibfnamefont {M.~T.}\
  \bibnamefont {Hansen}}, \bibinfo {author} {\bibfnamefont {R.~A.}\
  \bibnamefont {Brice{\~n}o}}, \bibinfo {author} {\bibfnamefont {R.~G.}\
  \bibnamefont {Edwards}}, \bibinfo {author} {\bibfnamefont {C.~E.}\
  \bibnamefont {Thomas}}, \ and\ \bibinfo {author} {\bibfnamefont {D.~J.}\
  \bibnamefont {Wilson}},\ }\href {\doibase 10.1103/PhysRevLett.126.012001}
  {\bibfield  {journal} {\bibinfo  {journal} {Phys. Rev. Lett.}\ }\textbf
  {\bibinfo {volume} {126}},\ \bibinfo {pages} {012001} (\bibinfo {year}
  {2021})}\BibitemShut {NoStop}%
\bibitem [{\citenamefont {K{\"o}rber}\ \emph {et~al.}(2020)\citenamefont
  {K{\"o}rber}, \citenamefont {Berkowitz},\ and\ \citenamefont
  {Luu}}]{Korber:2019cuq}%
  \BibitemOpen
  \bibfield  {author} {\bibinfo {author} {\bibfnamefont {C.}~\bibnamefont
  {K{\"o}rber}}, \bibinfo {author} {\bibfnamefont {E.}~\bibnamefont
  {Berkowitz}}, \ and\ \bibinfo {author} {\bibfnamefont {T.}~\bibnamefont
  {Luu}},\ }\href@noop {} {\bibfield  {journal} {\bibinfo  {journal}
  {arXiv:1912.04425 [hep-lat, physics:nucl-th, physics:physics]}\ } (\bibinfo
  {year} {2020})},\ \Eprint {http://arxiv.org/abs/1912.04425} {arXiv:1912.04425
  [hep-lat, physics:nucl-th, physics:physics]} \BibitemShut {NoStop}%
\bibitem [{\citenamefont {Leinweber}\ \emph {et~al.}(1991)\citenamefont
  {Leinweber}, \citenamefont {Woloshyn},\ and\ \citenamefont
  {Draper}}]{Leinweber:1990dv}%
  \BibitemOpen
  \bibfield  {author} {\bibinfo {author} {\bibfnamefont {D.~B.}\ \bibnamefont
  {Leinweber}}, \bibinfo {author} {\bibfnamefont {R.~M.}\ \bibnamefont
  {Woloshyn}}, \ and\ \bibinfo {author} {\bibfnamefont {T.}~\bibnamefont
  {Draper}},\ }\href {\doibase 10.1103/PhysRevD.43.1659} {\bibfield  {journal}
  {\bibinfo  {journal} {Phys. Rev. D}\ }\textbf {\bibinfo {volume} {43}},\
  \bibinfo {pages} {1659} (\bibinfo {year} {1991})}\BibitemShut {NoStop}%
\bibitem [{:20(2019)}]{:2019Parallelepiped}%
  \BibitemOpen
  \href@noop {} {\enquote {\bibinfo {title} {Parallelepiped},}\ }\bibinfo
  {howpublished}
  {\url{https://en.wikipedia.org/wiki/File:Special_cases_of_parallelepiped.svg}}
  (\bibinfo {year} {2019}),\ \bibinfo {note} {accessed: December
  2019}\BibitemShut {NoStop}%
\bibitem [{\citenamefont {{for the Hadron Spectrum Collaboration}}\ \emph
  {et~al.}(2015)\citenamefont {{for the Hadron Spectrum Collaboration}},
  \citenamefont {Wilson}, \citenamefont {Brice{\~n}o}, \citenamefont {Dudek},
  \citenamefont {Edwards},\ and\ \citenamefont {Thomas}}]{Wilson:2015dqa}%
  \BibitemOpen
  \bibfield  {author} {\bibinfo {author} {\bibnamefont {{for the Hadron
  Spectrum Collaboration}}}, \bibinfo {author} {\bibfnamefont {D.~J.}\
  \bibnamefont {Wilson}}, \bibinfo {author} {\bibfnamefont {R.~A.}\
  \bibnamefont {Brice{\~n}o}}, \bibinfo {author} {\bibfnamefont {J.~J.}\
  \bibnamefont {Dudek}}, \bibinfo {author} {\bibfnamefont {R.~G.}\ \bibnamefont
  {Edwards}}, \ and\ \bibinfo {author} {\bibfnamefont {C.~E.}\ \bibnamefont
  {Thomas}},\ }\href {\doibase 10.1103/PhysRevD.92.094502} {\bibfield
  {journal} {\bibinfo  {journal} {Phys. Rev. D}\ }\textbf {\bibinfo {volume}
  {92}},\ \bibinfo {pages} {094502} (\bibinfo {year} {2015})}\BibitemShut
  {NoStop}%
\bibitem [{\citenamefont {Briceno}\ \emph {et~al.}(2017)\citenamefont
  {Briceno}, \citenamefont {Dudek}, \citenamefont {Edwards},\ and\
  \citenamefont {Wilson}}]{Briceno:2016mjc}%
  \BibitemOpen
  \bibfield  {author} {\bibinfo {author} {\bibfnamefont {R.~A.}\ \bibnamefont
  {Briceno}}, \bibinfo {author} {\bibfnamefont {J.~J.}\ \bibnamefont {Dudek}},
  \bibinfo {author} {\bibfnamefont {R.~G.}\ \bibnamefont {Edwards}}, \ and\
  \bibinfo {author} {\bibfnamefont {D.~J.}\ \bibnamefont {Wilson}},\ }\href
  {\doibase 10.1103/PhysRevLett.118.022002} {\bibfield  {journal} {\bibinfo
  {journal} {Phys. Rev. Lett.}\ }\textbf {\bibinfo {volume} {118}},\ \bibinfo
  {pages} {022002} (\bibinfo {year} {2017})},\ \Eprint
  {http://arxiv.org/abs/1607.05900} {arXiv:1607.05900} \BibitemShut {NoStop}%
\bibitem [{\citenamefont {{for the Hadron Spectrum Collaboration}}\ \emph
  {et~al.}(2018)\citenamefont {{for the Hadron Spectrum Collaboration}},
  \citenamefont {Brice{\~n}o}, \citenamefont {Dudek}, \citenamefont {Edwards},\
  and\ \citenamefont {Wilson}}]{Briceno:2017qmb}%
  \BibitemOpen
  \bibfield  {author} {\bibinfo {author} {\bibnamefont {{for the Hadron
  Spectrum Collaboration}}}, \bibinfo {author} {\bibfnamefont {R.~A.}\
  \bibnamefont {Brice{\~n}o}}, \bibinfo {author} {\bibfnamefont {J.~J.}\
  \bibnamefont {Dudek}}, \bibinfo {author} {\bibfnamefont {R.~G.}\ \bibnamefont
  {Edwards}}, \ and\ \bibinfo {author} {\bibfnamefont {D.~J.}\ \bibnamefont
  {Wilson}},\ }\href {\doibase 10.1103/PhysRevD.97.054513} {\bibfield
  {journal} {\bibinfo  {journal} {Phys. Rev. D}\ }\textbf {\bibinfo {volume}
  {97}},\ \bibinfo {pages} {054513} (\bibinfo {year} {2018})}\BibitemShut
  {NoStop}%
\bibitem [{\citenamefont {Bernard}\ \emph {et~al.}(2008)\citenamefont
  {Bernard}, \citenamefont {Lage}, \citenamefont {Mei{\ss}ner},\ and\
  \citenamefont {Rusetsky}}]{Bernard:2008ax}%
  \BibitemOpen
  \bibfield  {author} {\bibinfo {author} {\bibfnamefont {V.}~\bibnamefont
  {Bernard}}, \bibinfo {author} {\bibfnamefont {M.}~\bibnamefont {Lage}},
  \bibinfo {author} {\bibfnamefont {U.-G.}\ \bibnamefont {Mei{\ss}ner}}, \ and\
  \bibinfo {author} {\bibfnamefont {A.}~\bibnamefont {Rusetsky}},\ }\href
  {\doibase 10.1088/1126-6708/2008/08/024} {\bibfield  {journal} {\bibinfo
  {journal} {J. High Energy Phys.}\ }\textbf {\bibinfo {volume} {2008}},\
  \bibinfo {pages} {024} (\bibinfo {year} {2008})},\ \Eprint
  {http://arxiv.org/abs/0806.4495} {arXiv:0806.4495} \BibitemShut {NoStop}%
\end{thebibliography}%

\end{document}